\documentclass[12pt, preprint]{aastex}
\usepackage{graphicx,epstopdf, rotate,amsmath}
\DeclareGraphicsExtensions{.ps}
\usepackage{lscape}
\slugcomment{To appear in the Astronomical Journal}

\shorttitle{Title}
\shortauthors{McCarthy and Wilhelm}

\begin{document}


\title{Characterizing the AB Doradus Moving Group Via High Resolution Spectroscopy and Kinematic Traceback}


\author{Kyle McCarthy and Ronald J. Wilhelm}
\affil{Department of Physics and Astronomy, University of 
Kentucky, Lexington, KY 40508-0055}

\email{kyle.mccarthy@uky.edu}


% \altaffiltext{1}{Visiting Astronomer, Cerro Tololo Inter-American Observatory.
% CTIO is operated by AURA, Inc.\ under contract to the National Science
% Foundation.}
% \altaffiltext{2}{Society of Fellows, Harvard University.}


\begin{abstract}
We present a detailed analysis of 10 proposed F and G members of the nearby, young moving group AB Doradus (ABD). Our sample was obtained using the 2.7m telescope at the McDonald Observatory with the coude echelle spectrograph, achieving R $\sim$ 60,000 and S/N $\sim$ 200.  We derive spectroscopic T$_{eff}$, log(g), [Fe/H], and microturbulance (v$_{t}$) using a bootstrap method of the TGVIT software resulting in typical errors of 33K in T$_{eff}$, 0.08 dex in log(g), 0.03 dex in [Fe/H], and 0.13 km s$^{-1}$ in v$_{t}$. Characterization of the ABD sample is performed in three ways: (1) Chemical homogeneity, (2) Kinematic Traceback, and (3) Isochrone fitting. We find the average metal abundance is [M/H] = -0.03 $\pm$ 0.06 with a traceback age of 125 Myrs. Our stars were fit to 3 different evolutionary models (Siess et al. 2000, Baraffe et al. 1998, and YREC) and we found the best match to our ABD sample is the YREC [M/H] = -0.1 model. In our sample of 10 stars, we identify 1 star which is a probable non-member, 3 enigmatic stars, and 6 stars with confirmed membership. We also present a list of chemically coherent stars from this study and the Barenfeld et al. (2013) study.
\end{abstract}


\keywords{Stars: fundamental parameters, Stars: pre-main sequence}

\section{Introduction}

The discovery of nearby, young moving groups has provided a fortuitous window into the processes and evolution of stars and planets at pre-main sequence (PMS) ages (see Zuckerman \& Song 2004 and Torres et al. 2008 for a review). These young stars allow the study of proto-planetary disks and their dissipation (France et al. 2011, Schneider et al. 2012, Brandt et al. 2014) in addition to the closely related topic of planet formation (Bryden et al. 2009, Biller et al. 2013, Rodigas et al. 2014). They also provide meaningful insight into lithium depletion rates (Mentuch et al. 2008, Yee \& Jenson 2010, Binks \& Jeffries 2013) and stellar activity (Scholz et al. 2007, Biazzo et al. 2007, Murgas et al. 2013). In the future, it may even be possible to detect planets around chromospherically active stars which will help constrain planet formation mechanisms (Moulds et al. 2013, Jeffers et al. 2014). The past decade has seen a wealth of research attempting to find new members of these young groups (Torres et al. 2006, L\'{e}pine \& Simon 2009, Schlieder et al. 2010, Malo et al. 2013, Rodriguez et al. 2013, Zuckerman et al. 2013) as well as classifying and analyzing general properties of currently proposed members (Biazzo et al. 2012, De Silva et al. 2013, Barenfeld et al. 2013, hereafter Ba13). 

Of the $\sim$10 moving groups within 100 pc, the most intensely studied is AB Doradus (ABD), named after the bright K star quadruple system. Upon its discovery, Zuckerman et al. (2004) believed ABD to have an age of nearly 50 Myr based on its lower H$\alpha$ emission and more evolved M-type stars compared to the younger Tucana/Horologium moving group (Tuc/Hor, 30 Myr); however, the next year Luhman et al. (2005) found an isochronal age much older than 50 Myr. This study used an M$_{K}$ versus $V - K$ diagram with all proposed members from Zuckerman et al. (2004) and found ABD was more evolved than the 35-50 Myr cluster IC 2391 and matched nearly identically with the Pleiades open cluster ($\sim$125 Myr). Ortega et al. (2007) also found this result by tracing back the orbits of both ABD and the Pleiades and found they were closest together at 125 Myr. The Ba13 study has taken this a step further and performed a kino-chemical comparison on individual members and found an age consistent with the previous Luhman et al. (2005) and Ortega et al. (2007) studies. This older age is beneficial for spectroscopic analysis since the F and G type stars have slowed their rotation rates, greatly reducing the severe line blending due to rotational broadening that would otherwise hamper a detailed analysis. 

The aim of this work is classifying F and G type stars from ABD to provide a better picture of the groups spectroscopic properties. We decided to test ABD due to its close proximity to the sun and northern declination. ABD also has the largest number of proposed members from the 9 moving groups listed in da Silva et al. (2009) with 187 potential members (127 from McCarthy \& White 2012 and references therein, 3 from Schlieder et al. 2012a, 8 from Schlieder et al. 2012b, 45 from Malo et al. 2013, 2 from Rodriguez et al. 2013, 2 from Riedel et al. 2014). Of these 187, there are 7 F, 31 G, and 10 K0/1 type stars.  

In Section 2 we discuss our sample and reduction techniques. Sections 3 and 4 describe the spectroscopic and chemical analysis while Section 5 shows the kinematic traceback for these stars. Section 6 discusses evolutionary models and placement on the HR Diagram. In Section 7 we make notes about each star and compare with Ba13 and conclude in Section 8.

\section{Observations and Data Reduction}

We obtained high resolution (R $\sim$ 60,000) echelle spectra of 10 ABD stream stars\footnote{Zuckerman et al. (2004) noted the ABD consists of a nucleus of 10 stars and a surrounding stream of young stars. The stars in our sample are all members of the stream.}. These observations were conducted using the 2.7 meter telescope at the McDonald Observatory with the coud\'{e} spectrograph in the TS23 operating mode providing wavelength coverage from $\sim$ 4,000 to 9,000 \AA\,. A semi-automated reduction procedure was performed using PyRAF\footnote{PyRAF is a product of the Space Telescope Science Institute, which is operated by AURA for NASA.} by first subtracting the bias frames and dividing flats, removing the scattered light, extracting the apertures, calibrating wavelengths with Th-Ar lamps and applying the appropriate dispersion. Continuum normalization was performed using the \emph{continuum} package in PyRAF (typically 3rd order polynomial).  No extinction corrections were applied due to the close proximity of these stars. The signal to noise ratio (SNR) is $\sim$ 200.

In addition to these 10 ABD stars, we observed 12 stars from the Valenti \& Fischer (2005; VF05) catalog (SPOCS) and the sun to ensure a reliable reduction and analytic process. VF05 have R $\sim$ 70,000 data which they passed through a grid of synthesized spectra and found parameters which minimize the $\chi^{2}$ between the observed data and the synthetic spectra. Throughout this work we will verify our procedure by checking for consistency between the values obtained using our methods with the values quoted in VF05.

\section{Spectroscopic Analysis}

\subsection{T$_{eff}$, log(g), v$_{t}$, [Fe/H]} 
\begin{figure}[t]
	\centering
		\includegraphics[scale=.7,angle=90]{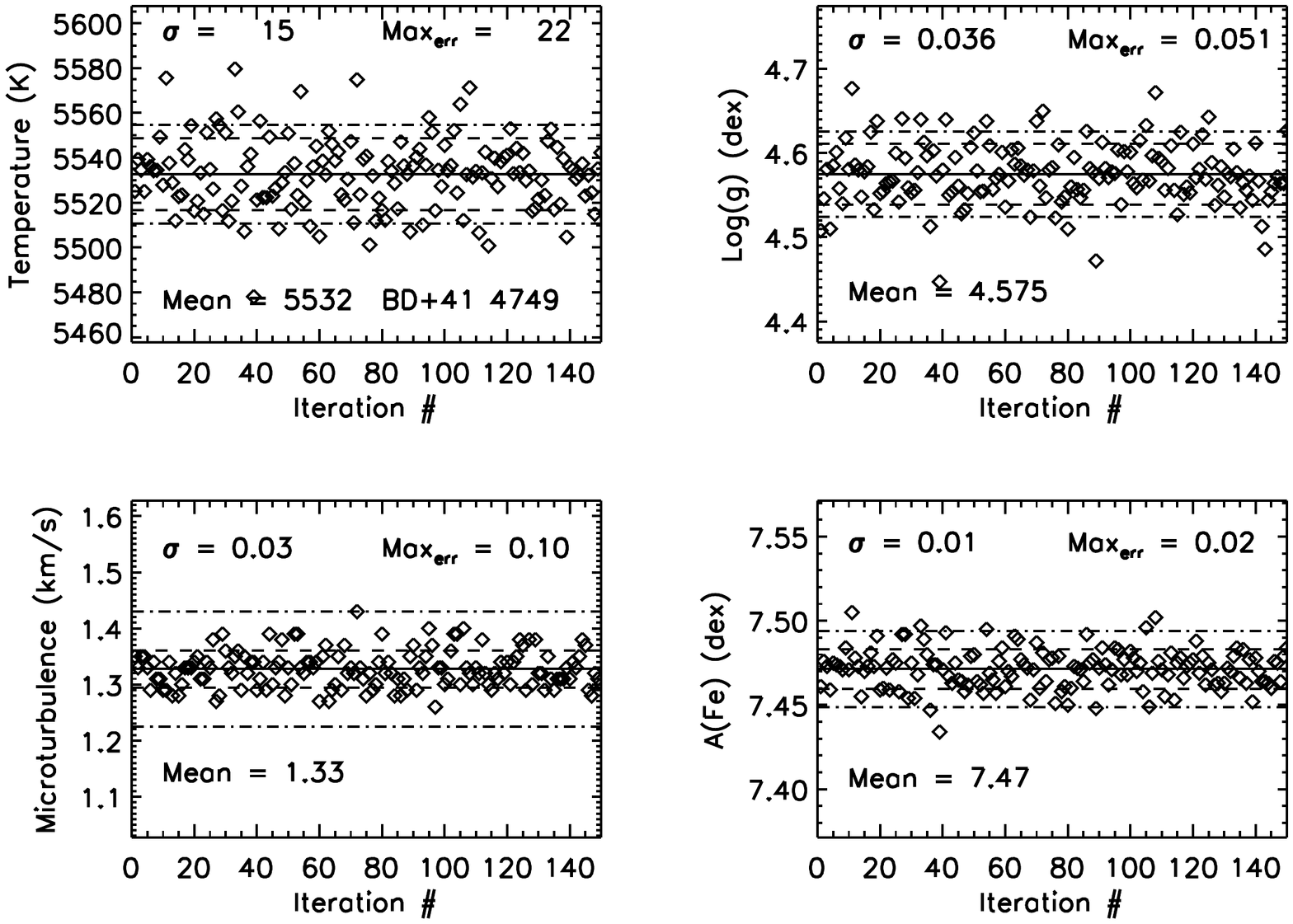}
	\caption{Results from the bootstrap method. The $\sigma$ uncertainties are from the standard deviation of results with the Max$_{err}$ is the $\sigma$ uncertainty added in quadrature to the uncertainties output from TGVIT.}
	\label{bootstrap}
\end{figure}

To measure T$_{eff}$, log(g), microtubluence (v$_{t}$), and [Fe/H]\footnote{[Fe/H] are derived using A(H) = 12 and A(Fe)$_{\odot}$ = 7.48 taken from our own solar spectrum} shown in Columns 3-6 of Table \ref{Param}, we start by using the line list, log(gf) values, and excitation potentials from Takeda et al. (2005). We measure Equivalent Widths (EWs) for 246 FeI and 22 FeII lines in IRAF\footnote{IRAF is distributed by the National Optical Astronomy Observatories, which are operated by the Association of Universities for Research in Astronomy, Inc., under cooperative agreement with the National Science Foundation.} and keep only lines where the EW is less than $\sim$ 100 m\AA\, as larger EWs are prone to errors in the dampening coefficient and enhanced contribution from the wings. We also remove lines that are below 5 m\AA\, due to noise contamination, lines which are affected by cosmic ray hits, or lines that are too badly blended to contribute believable EWs. For these reasons, no two stars have the same final line list, and the final line list for our moving group stars averaged 158 FeI lines and 12 FeII lines. We chose to use the TGVIT\footnote{http://optik2.mtk.nao.ac.jp/$\sim$takeda/tgv/} code (Takeda et al. 2002, Takeda et al. 2005) which uses the typical excitation and ionization balance in an iterative procedure to solve for the above parameters. To produce precise results, we use a bootstrap method by randomly selecting 90$\%$ of the FeI and FeII lines, executing TGVIT, running the automated \emph{checkup} tool which removes lines that differ by $>$2.5$\sigma$, and repeated this process four times to ensure no lines are driving a bad fit. We run 150 iterations for each star. The archetype of this procedure is shown in Figure \ref{bootstrap}. 

The errors involved in this method are from two main sources: the EWs and the iteration process in TGVIT. We estimate a generous 5$\%$ error in our EWs based on continuum placement and SNR. This error is implemented in our bootstrap code via a random error between 0 and 5\% on the input EW. TGVIT determines errors in the program by individually adjusting each parameter until one of the following conditions is not satisfied: abundance vs. excitation potential independence, abundance vs. equivalent width independence, or ionization balance. These errors are independent of each other and thus we add in quadrature the standard deviation of the bootstrap method to the average error found in TGVIT and report them in Table \ref{Param}. After a detailed analysis, it was found that covariance errors were negligible since TGVIT varies temperature and surface gravity simultaneously. Full treatment can be seen in Appendix A.

In addition to the above technique, we also performed an analysis using the 2013 version of MOOG\footnote{http://www.as.utexas.edu/$\sim$chris/moog.html} (Sneden 1973) using the same line list and atomic parameters as were used in the TGVIT analysis. Our procedure is as such: (1) Run MOOG under the abfind setting, utilizing interpolated Kurucz ATLAS9 models (Castelli \& Kurucz 2004) with no convective overshoot and an initial estimate of the parameters given by TGVIT; (2) Clip features whose abundance varies by 2.5$\sigma$ and iterate until no lines are above this threshold; (3) Adjust each parameter until achieving the minimum slope in Abundance of FeI (A(FeI)) vs. Excitation Potential to find T$_{eff}$ and A(FeI) vs. Reduced Equivalent Width to find v$_{t}$, and achieve ionization balance between A(FeI) and A(FeII) to find log(g) and [Fe/H]. 

\begin{figure}[t]
	\includegraphics[scale=.7,angle=90]{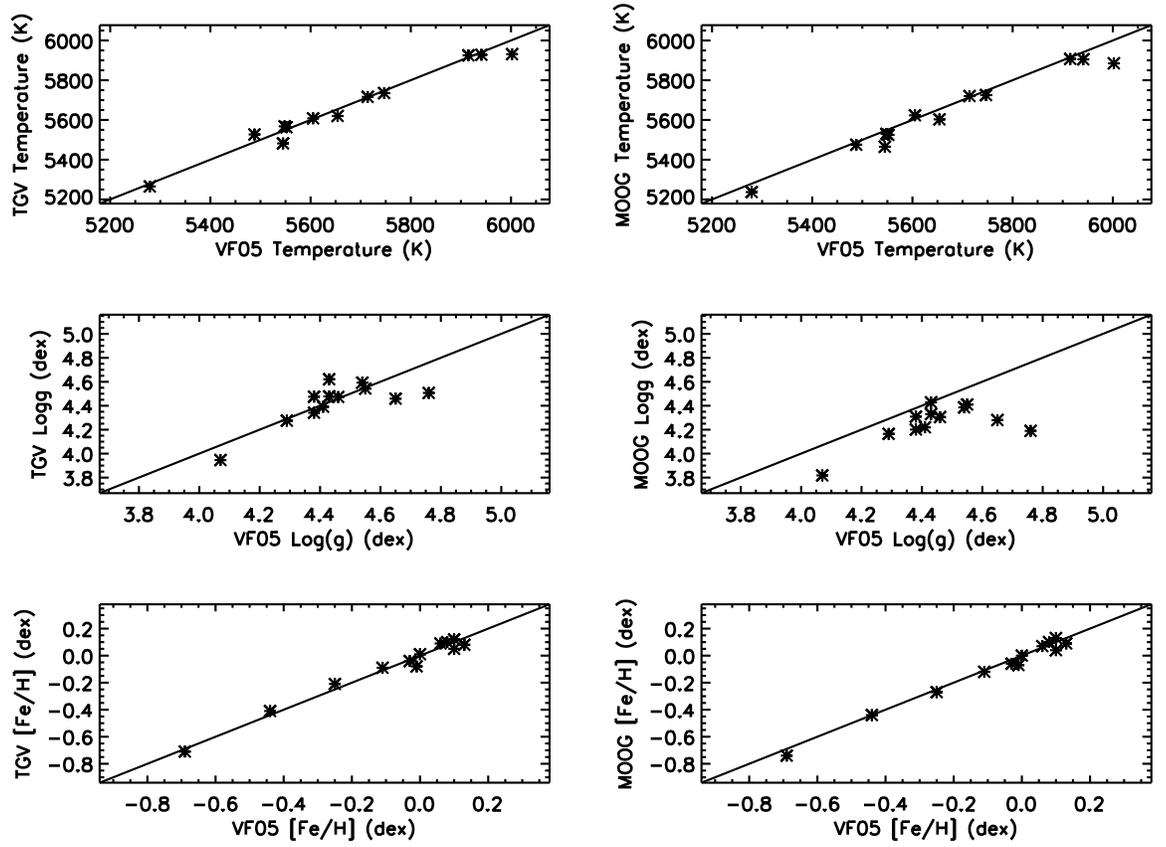}
	\caption{Comparison between VF05 values and those obtained in this work from TGVIT and MOOG. The temperature and abundance values are nearly identical between TGVIT and MOOG, however MOOG is roughly 0.15 dex lower in surface gravity.}
	\label{VF_Comp}
\end{figure}

Figure \ref{VF_Comp} display the results of both techniques performed on the VF05 sample. These results indicate that MOOG typically gives slightly smaller values for T$_{eff}$ and [Fe/H] and significantly smaller values for log(g). Lower surface gravities using MOOG are also found in Tsantaki et al. (2013) and Sousa et al. (2008) who compare surface gravities from MOOG to those derived from HIPPARCOS parallaxes (van Leeuwen 2007) and find that MOOG underestimates surface gravities by as much as 0.5 dex near log(g) $\sim$ 4.5. We therefore do not use the MOOG values and only report those found using the TGVIT bootstrap method; however, we would like to note that agreement with VF05 does not imply correctness. To further evaluated MOOG vs. TGVIT surface gravities, we test masses derived using both methods to masses derived using Torres et al. (2010) with corrections from Santos et al. (2013) in Section 3.2.

\subsection{Radius and Mass} 
Once these initial parameters are found, we use the prescription in McCarthy \& White (2012) to solve for the radius. All stars are found in the Tycho-2 Catalog (H$\o$g et al. 2000), giving Tycho $B_{T}$ and $V_{T}$ values which are converted into V magnitudes using the method in Bessell (2000). All but two stars (BD-04 1063 and BD-09 1034) have updated Hipparcos parallaxes (van Leeuwen 2007); for the other two, we use the distances listed in Torres et al. (2008) which are derived from a convergence method outlined in Torres et al. (2006). Bolometric corrections for the V band (BC$_{V}$) are found using the coefficients in Table 1 of Torres (2010). When applied to our solar spectrum, we find BC$_{V,\odot}$ = -0.07, corresponding to a solar bolometric magnitude of M$_{Bol,\odot}$ = 4.74 (Flower 1996). All of this information produced the luminosities for each star. The luminosities coupled with our computed temperatures are used to determine the final radius values. 

\begin{figure}[t]
	\includegraphics[scale=.7,angle=90]{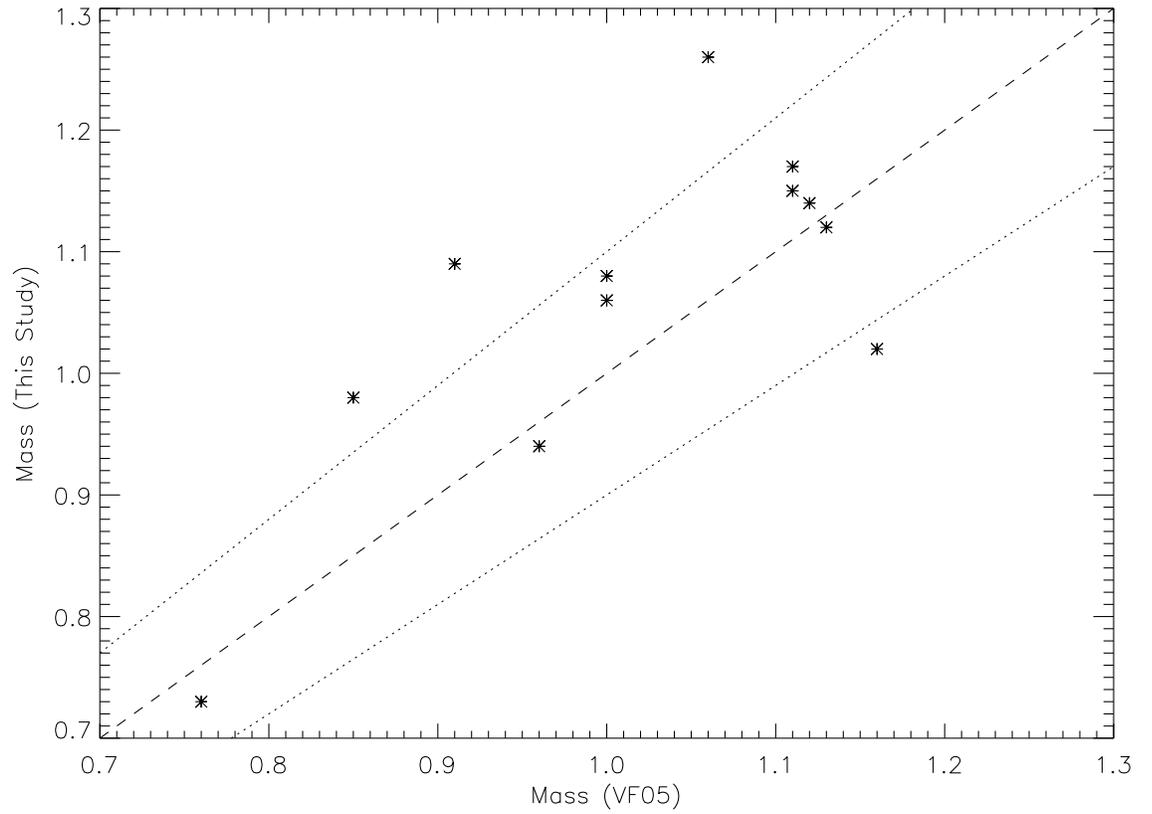}
	\caption{Comparison of masses between spectroscopically derived mass and masses from VF05 derived from evolutionary models. The dashed line shows a 1-to-1 correlation with the dotted line representing 10$\%$ errors.}
	\label{MassComp}
\end{figure}

We calculate the mass for each star using $M = \frac{g R^{2}}{G}$ where M is the mass of the star, g is the surface gravity, R is radius, and G is the gravitational constant. This method has been tested using main sequence evolutionary models in VF05 and Sousa et al. (2011). These studies found that masses derived from spectroscopic features are typically overestimated in when M $>$ 1 M$_{\odot}$ and underestimated when M $<$ 1 M$_{\odot}$. In Figure \ref{MassComp} we compare the masses found from the 12 VF05 stars using the spectroscopic technique in Section 3.1 with VF05 masses derived from evolutionary models. We find a 6$\%$ scatter between our values and VF05 with a minor offset of +0.05 M$_{\odot}$. This scatter is nearly identical to the average internal error in our derived VF05 masses of 0.06 M$_{\odot}$; therefore, we believe the bootstrap TGVIT method gives better estimates of the surface gravity which is the driving parameter in mass determination.

To further evaluate our mass estimates, we compared our masses for the VF05 stars to masses derived following Torres et al. (2010) with corrections made by Santos et al. (2013). After removing the giants from our sample, we found our masses were 0.01 $\pm$ 0.09 M$_{\odot}$ larger than the Torres et al. (2010) study, similar to the comparison against the VF05 evolutionary model masses. When applying the correction from Santos et al. (2013), we found a difference of 0.09 $\pm$ 0.09 M$_{\odot}$, still in agreement with our previous comparison. In addition to the TGVIT bootstrap values, we also found masses using the MOOG atmospheric parameters. Here, we found a -0.29 $\pm$ 0.13 M$_{\odot}$ difference between the masses estimated from MOOG parameters and the Torres et al. (2010) calibration. When applying the Santos et al. (2013) correction, masses still differed by -0.21 $\pm$ 0.13 M$_{\odot}$. We therefore find the masses using the TGVIT parameters to be more physically accurate than those using MOOG parameters.

Errors listed for the radii and masses in Table \ref{Param} are from four sources: the Tycho-2 colors, parallax estimates, T$_{eff}$ and log(g). We use standard error propagation formalism to find the final uncertainties. The first three sources of error primarily manifest themselves in the bolometric magnitude uncertainties which directly correspond to luminosity errors. These luminosity errors are the dominate source of error in the radius. Surface gravity errors are the dominate source of error in the mass.

\subsection{RV, vsin(i), and Lithium} 
\begin{figure}
	\centering
		\includegraphics[scale=.7,angle=90]{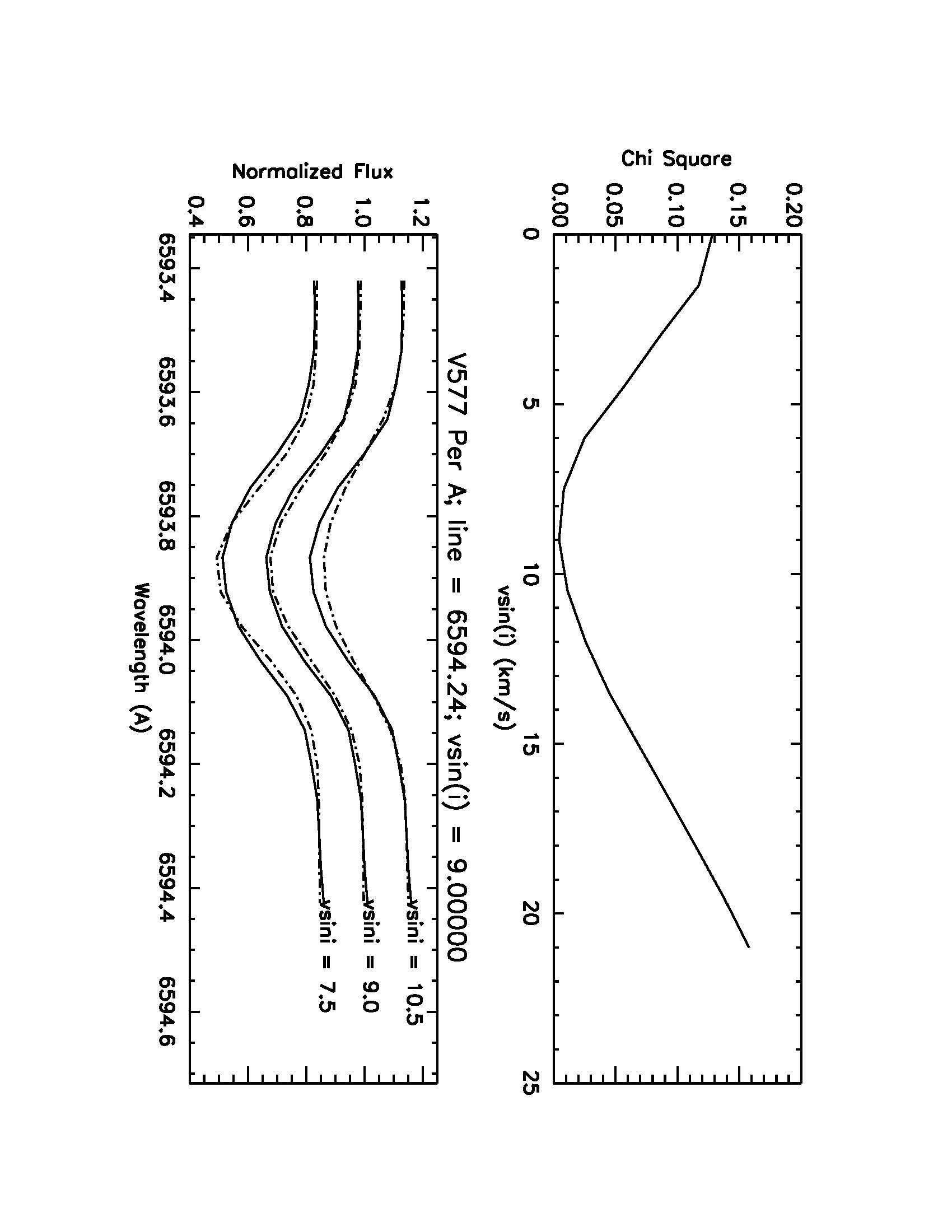}
		\caption{Tool for measuring vsin(i). The top panel shows the $\chi^{2}$ for each vsin(i) step and the bottom panel shows the best fitting model (dashed line) compared to the observed data (solid line) in the middle with one step slower on the bottom and one step faster on the top.}
	\label{vsini}
\end{figure}

Before the procedure in Section 3.1 was implemented, we measured the radial velocity (RV) of each star to determine the observed wavelength for each iron line. RVs were calculated by TAME (Kang \& Lee 2012), an automated EW estimator which allows the user to adjust the continuum level interactively. Occasionally blending will be so bad that TAME is not able to pick out the correct line. We take this into account by running an IDL procedure to remove lines whose RV's are greater than 2.5$\sigma$ away from the median value of all Fe lines. We then calculate the ratio of the EW found from TAME to the EW of the sun (found in IRAF) and plot this against the excitation potential of the line. A linear trend is fit to remove the temperature dependence and we select only lines which fall within 2.5$\sigma$ of this slope. The final RV is taken from the remaining lines. The average standard deviation between our method and VF05 is 0.8 km s$^{-1}$ and our radial velocities are slightly larger with an average difference of +0.6 km s$^{-1}$.

vsin(i) values are estimated using a $\chi^{2}$ method comparing the observed data against interpolated model spectra from the Kurucz ATLAS9 model atmospheres and using SPECTRUM (Gray \& Corbally 1994) to produce a high resolution synthetic spectrum. The synthetic spectra were degraded in resolution to match the FWHM of the ThAr emission features for each echelle order and a vsin(i) profile was convolved with spectrum as outlined by Gray (1992) in steps of 1.5 km s$^{-1}$. We found the minimum $\chi^{2}$ for 10 isolated FeI lines and averaged to find the final vsin(i) for the star. Figure \ref{vsini} presents this method where the top panel shows the $\chi^{2}$ for the particular line and the bottom panel displays the observed data in a solid line with the synthetic data in a dashed line. The middle spectral line is the best fit to the data. 

Occasionally, an observed feature will have a cosmic ray in its spectrum. This results in a very poor estimate of the feature and can skew the final result considerably. We account for this in the final stage by removing any feature whose vsin(i) is greater than 2$\sigma$ deviant from the average. Our errors are the 1$\sigma$ deviation in vsin(i) measurements.

Lithium EWs are found by measuring the $^{7}$Li 6708 line in IRAF. We estimate the same 5\% error on our measurements here as we did in Section 3.1.

\section{Chemical Analysis}

Nearby young moving groups provide an excellent testbed for the chemical homogeneity of a molecular cloud. It is presumed that each star in a moving group formed from the same parent molecular cloud, and thus the differences in elemental abundance can relay meaningful insights on how well each element is mixed. These abundances also provide information regarding potential planet hosting stars. Fischer \& Valenti (2005) showed that higher metalicity stars are more likely to harbor planets while more recent studies by Kang et al. (2011) and Adibekyan et al. (2012a) investigate specific abundance differences between known planet hosting stars and non-planet hosting stars. Kang et al. (2011) found an over-abundance of Mn in solar [Fe/H] planet hosting stars, though this result was not confirmed in Adibekyan et al. (2012a) who found the largest difference in Mg in the solar [Fe/H] regime (all-be-it a small difference). From these results, it is likely not possible to determine whether a star in ABD harbors a planet based solely on its chemical composition.

The elemental abundances for our ABD sample and VF05 stars are listed in Tables \ref{Abun} and \ref{Ctd}. Excitation potentials and oscillator strengths for all elements were from the Neves et al. (2009) study with corrections made in Adibekyan et al. (2012b). We further removed 2 SiI lines (5777.15\AA, 6527.21\AA), 1 CaI line (5867.56\AA), 2 TiI lines (4562.63\AA, 5145.47\AA), 1 V line (6081.45\AA), 4 CrI lines (4575.11\AA, 4600.75\AA, 6661.08\AA, 6882.52\AA), 1 Mn line (4502.21\AA), and 6 NiI lines (5010.94\AA, 5462.5\AA, 6175.37\AA, 6176.82\AA, 6177.25\AA, 6186.72\AA) due to lines being in between echelle orders or a bad feature in our solar spectrum. EWs for all lines are measured in IRAF and fed into MOOG under the \emph{abfind} setting with interpolated ATLAS9 model atmospheres. We perform this analysis on the solar spectrum taken during our observing run and the final [X/H] are the averaged line-to-line abundance differences between the star and the sun. 

Errors in [X/H] are derived similarly to Adibekyan et al. (2012b). The internal scatter is $\sigma$/$\sqrt{N}$ where N is the number lines and $\sigma$ is the standard deviation. We also include errors in the atmospheric parameters of the star given in Section 3.1 via a max/min method. Typical [X/H] uncertainties caused by the atmospheric parameters are $\pm$ 0.03 for temperature, $\pm$ 0.007 for logg, $\pm$ 0.03 for v$_{t}$, and $\pm$ 0.007 for [Fe/H]. Errors from the line-to-line scatter were added in quadrature to the errors caused by the atmospheric parameters. When only one line was available, the uncertainties given are 0.10 dex. 

\begin{figure}[h]
	\centering
		\includegraphics[scale=.7,angle=90]{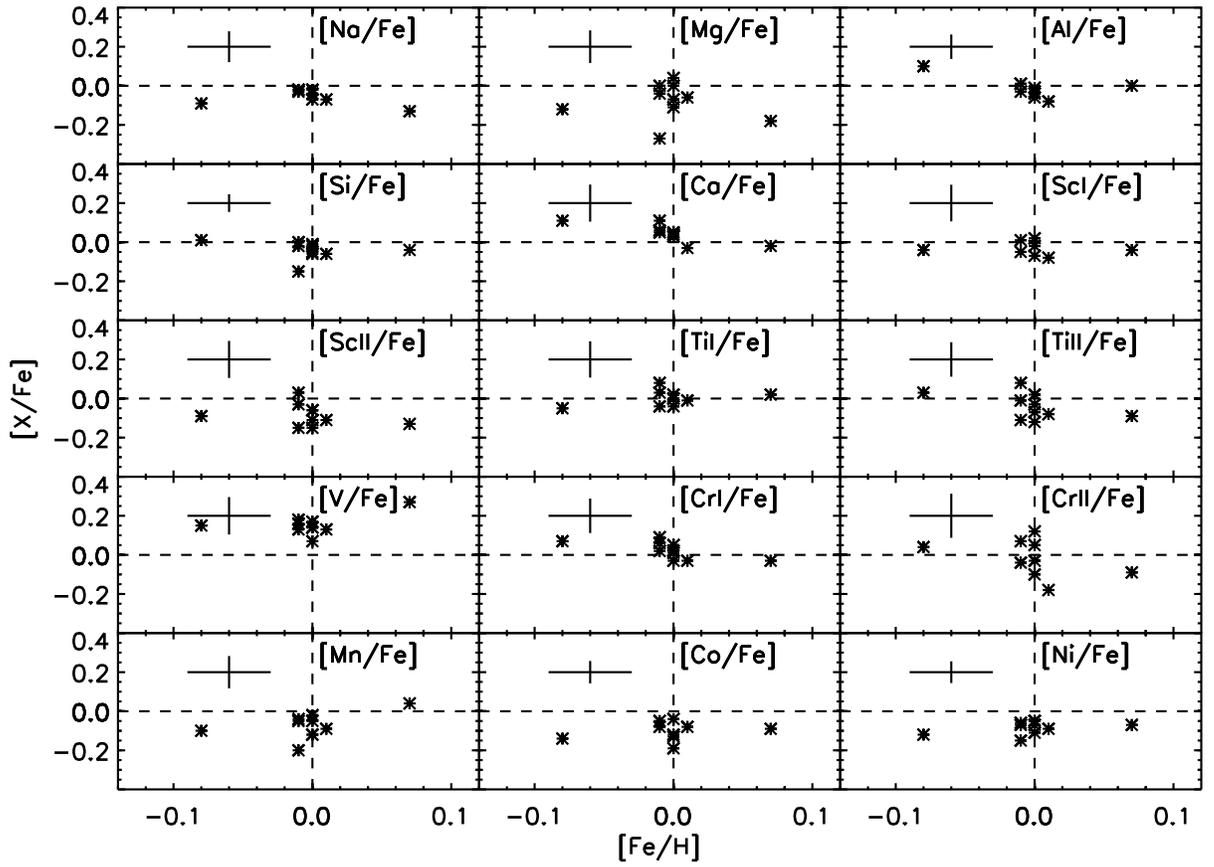}
		\caption{Metal abundances for 10 ABD stars. Typical errors are shown in the top left of each plot.}
	\label{ABDorMetals}
\end{figure}

Of the ABD sample listed in Tables \ref{Abun} and \ref{Ctd}, there are 2 stars with noticeable abundance differences. The first is BD-04 1063 which has the lowest Fe abundance in the sample. This star has significantly lower [Na/H], [TiI/H] and [Ni/H], while having slightly larger [Al/H]. In Figure \ref{ABDorMetals}, this star is typically equal to or above trends in [X/Fe] vs. [Fe/H]. It is possible that our measured Fe abundance is too low star for this star, and that if the abundance were solar, this star would follow most of the trends with the exception of Na, TiI, Mn, and Ni. The other outlier is BD-15 200 which, conversely, has the largest Fe abundance. While it remains consistent with the usual trends in [X/Fe], it does have a noticeably lower Mg abundance and larger Mn abundance. BD-15 200 (HD 6569) was also observed in Ba13 and they found similar results with moderate exceptions in Al, Si, and Mn and a large exception in Cr. Interestingly, in the Ba13 study, BD-15 200 was barely able to be claimed as chemically coherent with two other ABD stars which raises a flag about its membership in the group. 

In Figure \ref{ABDorMetals} we display the ABD sample of elemental abundances. The error bars are the average standard deviation given by MOOG. This plot displays the trends in [X/Fe] vs. [Fe/H]. The two stars mentioned earlier (BD-04 1063 and BD-15 200) are significant outliers in [Fe/H] and possible outliers in other elements.  If we assume each element has an equal weight in the overall metal abundance of a star, we calculate $<[M/H]>$ = -0.03 $\pm$ 0.06 both including and excluding BD-04 and BD-15. These results are similar to those found in Ba13 who found $<[M/H]>$ = 0.01 $\pm$ 0.02, with their study not including V or Mn but including Ba. In the Biazzo et al. (2012) study of 5 ABD stars, they did not investigate V or Mn, but did investigate Zn. While this study did not report [M/H] for the cluster, we calculated this value by converting the [X/Fe] from Table 4 of that paper to [X/H] and found $<[M/H]>$ = 0.06 $\pm$ 0.06, slightly more abundant compared to our study. Based on the similarity of these three analyses, it is likely that the average elemental abundances are consistent for all ABD stars. Therefore, if a star is found to not be chemically consistent with these results it should raise questions about its membership (such as BD-04 1065 and BD-15 200). 

\begin{figure}[b]
	\centering
		\includegraphics[scale=.7,angle=90]{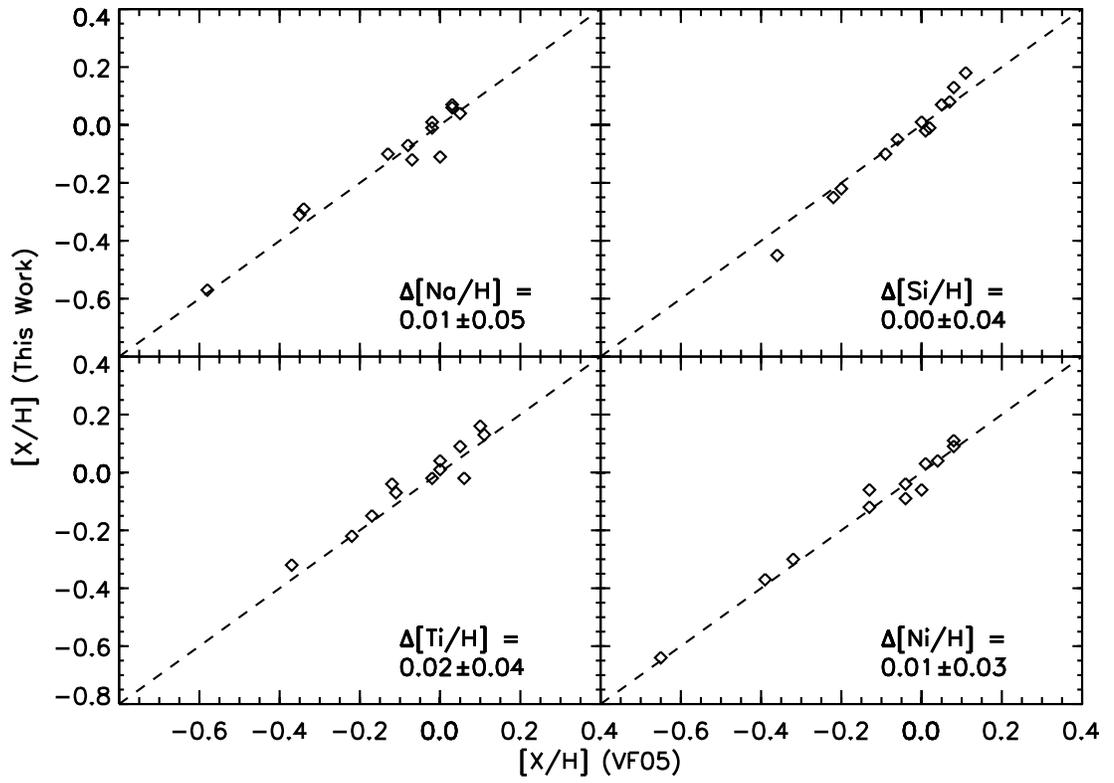}
		\caption{Comparison of parameters measured in this study to those given in VF05.}
	\label{VFMetals}
\end{figure}

Figure \ref{VFMetals} shows a comparison of our metal abundances vs. those found in VF05 to ensure our procedure is working properly. The VF05 study only found the abundance for Na, Si, Ti, Fe, and Ni, therefore we consider this a ``spot check'' of all metal abundances. The average difference between our values and VF05 regardless of the element is $\Delta$ [X/H] = 0.01 $\pm$ 0.04, where the error comes from the standard deviation. It is no surprise the best agreement is found in Ni as this element has $\sim$ 35 measurable spectral features and the worst agreement is in Na since there are only 3 lines. 

\section{Kinematic Traceback}

In addition to chemical homogeneity, stars of the same moving group should have a common origin. Estimating the inception of a moving group involves mapping a galactic potential and using current UVW velocities and XYZ positions to trace the motion back in time. This method is commonly referred to as a Kinematic Traceback (KTb). To traceback ABD, we use the equations of motion outlined by Asiain et al. (1999) in Section 2 of that paper. In brief, it is assumed that the galactic potential only consists of the halo, bulge, and disk components with no contribution from other potentials such as spiral arms or the bar, nor do we consider effects cause by heating processes. We use the heliocentric coordinates $\xi'$, $\eta'$, and $\zeta'$ which are co-moving with the sun where $\xi'$ is positive toward the galactic anti-center, $\eta'$ is positive in the direction of the suns orbit, and $\zeta'$ is positive coming out of the galactic plane. The epicyclic approximation is used over more complex treatment as these stars are in nearly circular orbits and do not have large peculiar velocities. The equations of motion are found from Equations (1) and (2) of Asiain et al. (1999). For our calculations, the epicyclic frequency and vertical frequency are roughly 40.0 km s$^{-1}$kpc$^{-1}$ and 73.5 km s$^{-1}$kpc$^{-1}$ respectively. The galactic variables used in our KTb code are found in Table \ref{Gal}.

Table \ref{UVW} displays the initial positions and velocities of our stars. Using parallaxes and RA/DEC coordinates from Hipparcos and RVs from this work, UVW space motions are found using the publicly available \emph{gal\textunderscore uvw} IDL program and xyz positions are calculated from a self produced IDL routine. For BD-04 1063 and BD-09 1034, the two stars not listed in Hipparcos, we use distances from Torres et al. (2008) and RA/DEC coordinates from the Tycho-2 catalog.  

\begin{figure}[h]
	%\centering
		\includegraphics[scale=.7,angle=90]{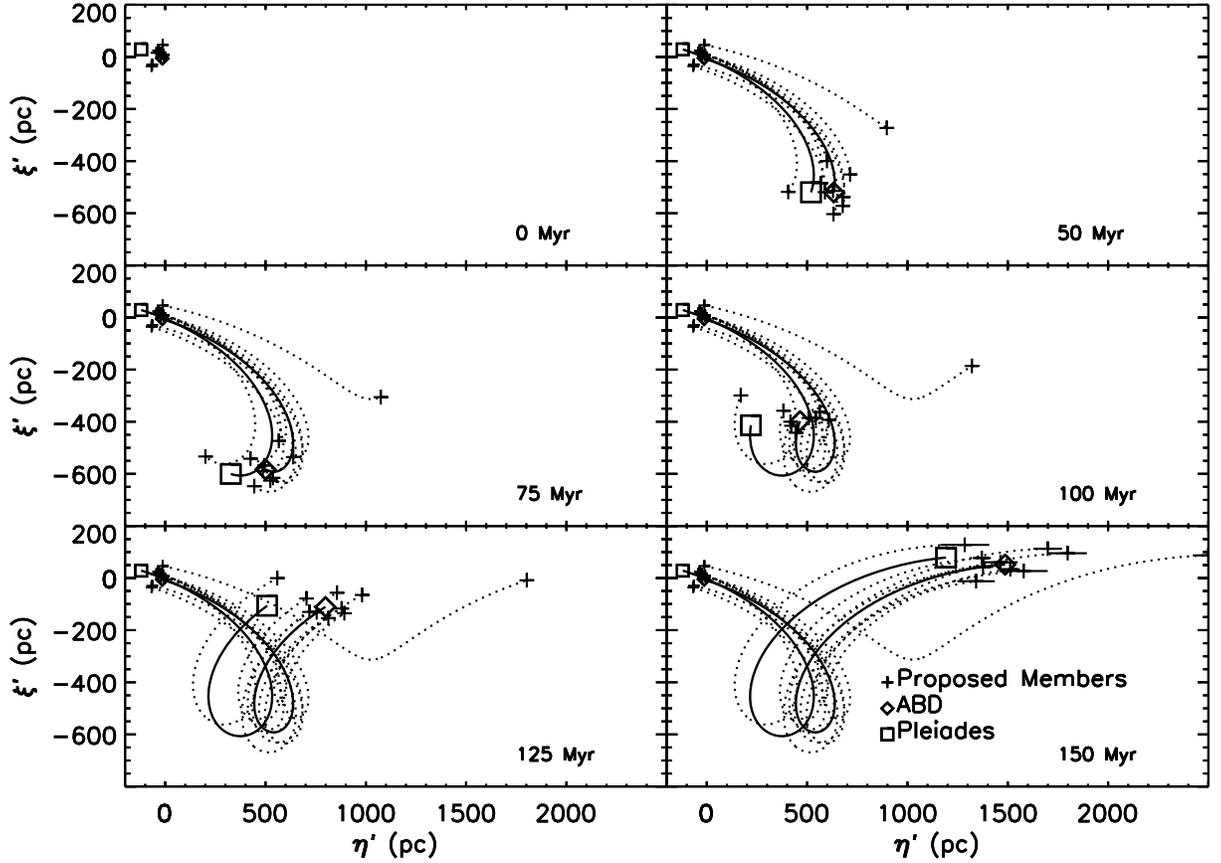}
		
	\caption{Kinematic Traceback of the 10 ABD stars (plus signs) along with bulk properties of ABD (Diamond) and the Pleiades open cluster (square). The outlying stars from left to right are BD+37 604A, BD-15 200, and BD+41 4749. The seven stars which occupy the same region in the 125 Myr plot we deem as kinematically consistent set of ABD stars.}
	\label{ABDtrace}
\end{figure}

In Figure \ref{ABDtrace}, we show the KTb of stars in our ABD moving group sample as well as the average motion of the Pleiades open cluster (square symbol) and ABD (diamond symbol) moving groups. We follow the procedure outlined in Ortega et al. (2007) to find the present velocity and positions for the Pleiades and use the values listed in Torres et al. (2008) for ABD. Errors present are from Hipparcos data and our RVs. We run 1,000 traceback realizations including random, gaussian distributed errors and display the average position of each star at the given time step. In the bottom right panel (150 Myr) the size of the points represent the standard deviation among final positions the 1,000 trajectories; typical errors are $\sim \pm$ 75 pc. Likewise, the errors listed in Table \ref{UVW} are the standard deviation of the 1,000 initial positions and velocities for each star.

From Figure \ref{ABDtrace}, there are 8 stars which share similar tracebacks and 7 stars which occupy the same location at 125 Myr. These 7 stars are kinematically consistent with the general motion of ABD and were likely formed at the same time; therefore, we consider this the ``locus'' of our sample. In the 125 Myr panel, the outlying stars from left to right are BD+37 604 A, BD-15 200, and BD+41 4749. BD+37 604 A has a slightly more negative U velocity and RV as well as slightly smaller V and W velocities than the bulk of ABD, likely attributing to its deviant motion. Interestingly, this star appears to follow more closely to the Pleiades than to ABD. The star which follows a completely separate trajectory from ABD is BD+41 4749. This is likely due to its large negative radial velocity and smaller U velocity. Though this star is wildly inconsistent with the other ABD members in the traceback, its surface gravity and lithium abundance show its youth and therefore we do not dismiss this star from the group based on one test; however, it does raise questions about its membership.

\begin{figure}[h]
	%\centering
		\includegraphics[scale=.7,angle=90]{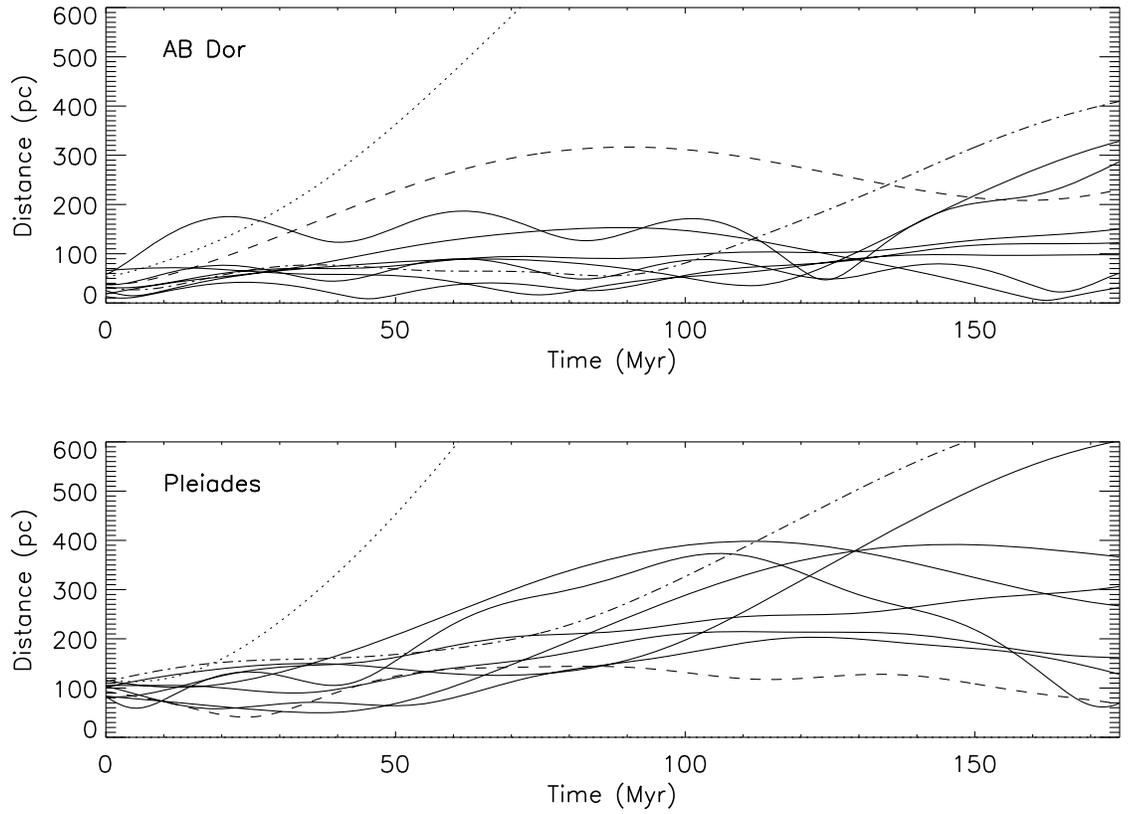}
		
	\caption{Distances of individual stars to ABD and the Pleiades open cluster. 7 of the ABD members converge just after the 125 Myr with the outlying stars being BD+47 4749 (dotted), BD+37 604 A (dashed), and BD-15 200 (dot-dashed).}
	\label{Flyby}
\end{figure}

In addition to tracing the orbits of individual stars, Figure \ref{Flyby} displays the distance from each star to the bulk motion of ABD (top) and the Pleiades (bottom). Our sample diverges rapidly from the Pleiades, implying ABD is its own separate group. In the top panel, there are 7 stars which converge to less than 100 pc from ABD between 123-130 Myr, the exact time scale predicted by Luhman et al. (2005), Ortega et al. (2007), and Ba13. This convergence has been refereed to as the ``focusing phenomena'' in Yuan (1977) and Yuan \& Waxman (1977). While this data is suggestive of a $\sim$125 Myr group, Soderblom (2010) notes a fundamental limitation on kinematic ages of 20-30 Myr due to interactions with massive objects (e.g. molecular clouds or other stars) as stars go through their orbits. We therefore proceed with caution when making claims of the age of ABD using the KTb. Additionally, the divergent stars could be remnants of a perturbation, giving rise to their strange motions.

Of the three most divergent stars at 125 Myr in Figure \ref{Flyby}, BD+37 604 A (dashed line) and BD+41 4749 (dotted line) were listed above and the third is BD-15 200 (dot-dashed line). This star remains close to ABD for 100 Myrs before dramatically deviating from the rest of the group. As noted in Section 4, this star has the largest metal abundance in our sample. One conjecture for its deviation at 100 Myrs is this star formed at a later epoch of star formation, after the most massive stars enriched the environment with more metals. This type of age separation has been observed by Palla et al. (2007) in the Orion Nebula Cluster, finding an older population (10-30 Myr) along with a younger population (1-2 Myr), implying a time-dependent pattern of star formation.

\section{Comparison to Evolutionary Models}

Often times directly determining the mass or radius of a star is not possible as the star may not have a companion or the separation of its companion is too far to measure accurate orbital parameters. Heavy emphasis is then placed on stellar evolutionary models to estimate the mass and radius. Main sequence models have been empirically refined thanks to long baseline interferometers such as the CHARA Array (ten Brummelaar et al. 2005) which can resolve stellar radii down to $\sim$ 0.34 milli-arcseconds (Baines et al. 2012), enough to resolve the radius of individual stars or determine orbital elements of nearby binary stars (see Raghavan et al. 2009, White et al. 2013). However, in the PMS regime there are upwards of 10 evolutionary models, all with different input physics leading to systematic differences in radius and mass between 50-200 \% depending on the choice of model (Hillenbrand \& White 2004). Using nearby, young moving groups is pivotal for constraining PMS models since we can accurately determine temperatures and luminosities, place the stars on an HRD and thus provide an empirical test of the models. This way, when the PMS models are applied to fainter groups, the only uncertainties will be from observations, not from the models.

To begin this empirical test, in Figure \ref{Iso_Paper} we display evolutionary models from 3 prominent groups, Siess et al. (2000, SDF00) in the top left panel, Baraffe et al. (1998, BCAH98) in the top right, and YREC\footnote{The isochrones are publically available at http://www.astronomy.ohio-state.edu/iso/} (in prep) in the bottom two panels. Our sample of 10 ABD stars is plotted on top of these models. There is a thin isochrone of 8 ABD stars around the 100 Myr, consistent with previous estimates for the moving groups age (see Ortega et al. 2007, Ba13). Two stars lie above this isochrone around 60 Myr. This elevation could be caused by errors in the parameters of these stars, formation in a later epoch, or potentially not being associated with the moving group entirely.

The latter hypothesis is likely for the hottest outlying star, BD-04 1063, whose age is nearly 60 Myr. This star is also the most metal poor of our sample and it is possible that it could be a member of a slightly younger, less metal rich moving group. The other outlying star is BD+23 296 A who has a K4 companion at $1\farcs8$ (Torres et al. 2008). While the Tycho-2 catalog can resolve separations of $\sim 0\farcs8$, it could be possible that its companion or a potentially unresolved companion and is adding flux into the V band which would lift the star above the isochrone. It could also be that the parallax is incorrect. The original Hipparcos parallax is 30.99 milli-arcseconds (mas) (Perryman \& ESA 1997); however, the updated parallax in van Leeuwen (2007) places this star 4 pc farther away with a parallax of 27.3 mas which consequently leads to a larger luminosity. Placing the star 4 pc closer has a profound effect on the luminosity, but it does not completely rectify the problem. In order for BD+23 296 A to fall along the same isochrone as ABD, it would have to be 10 pc closer than the van Leeuwen (2007) distance, therefore it may be a combination of incorrect parallaxes and additional binary flux that leads to this stars elevation on the HRD.

In addition to moving group membership, these plots also give important information regarding the evolutionary models themselves. The thin 100 Myr isochrone places these stars on the bottom of the main sequence and provides a firm lower limit on where the PMS models should reach. It is clear that neither the SDF00 nor BCAH98 models accurately predict the transition from PMS to ZAMS. The models which best fit the data are those from YREC. While the [M/H] = 0.0 models empirically match the ABD data, the [M/H] = -0.1 provide the best fit to the data. The best fit was determined by finding the mean minimum distance from each star (exculding the two above) to the 100 Myr isochrone on the HRD. This result is generally consistent with our findings in Section 4 that the group is slightly less metal abundant than the sun. At the moment, the YREC models most accurately describe the transition from PMS to ZAMS. In order to make further claims, several moving group with different ages should be tested against these models to see how well each model matches the empirical data.

\section{Notes on Individual Stars and Comparison to Ba13}

\begin{figure}
	\centering
		\includegraphics[scale=.7,angle=90]{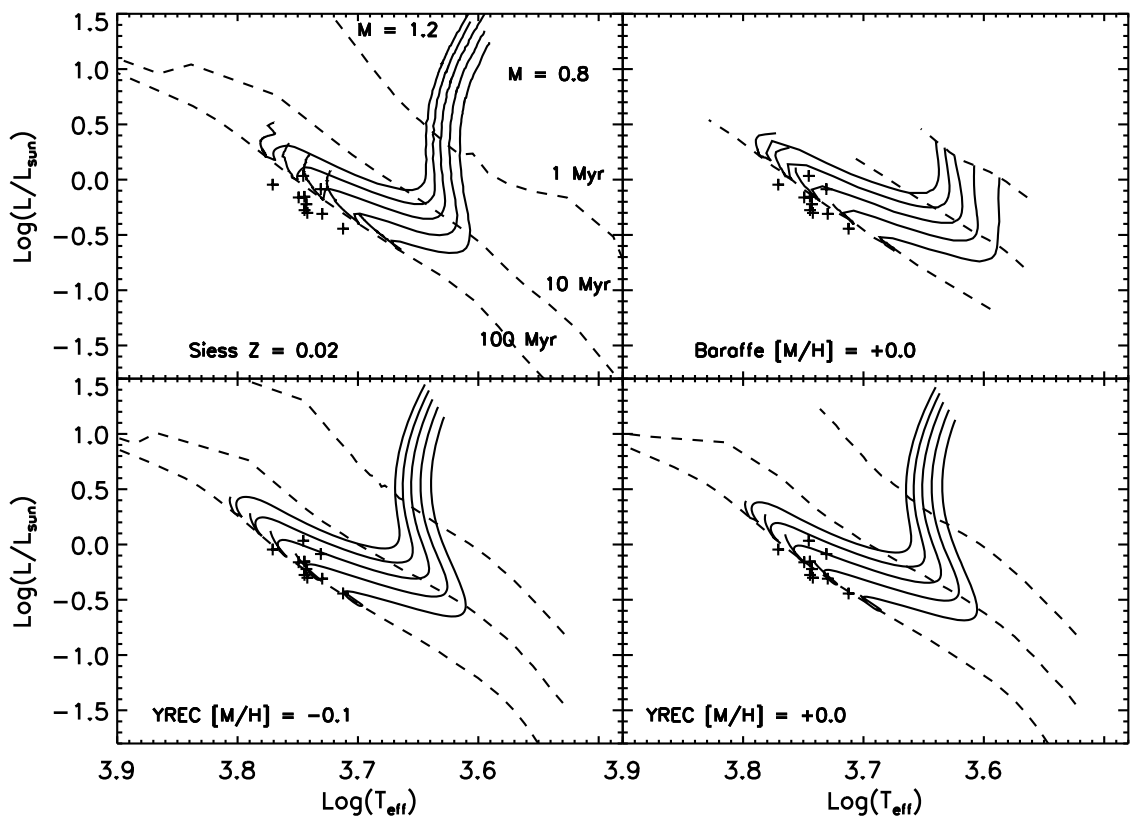}
		\caption{Comparison of three stellar evolutionary models to the observed data. The models shown are SDF00 (top left), BCAH98 (top right) and YREC ([M/H] = -0.1 on left, [M/H] = 0.0 on right). The YREC [M/H] = 0.0 models show the best empirical fit to our data. Error bars are roughly the size of the symbols.}
	\label{Iso_Paper}
\end{figure}

\emph{BD-04 1063}: This star is peculiar in nearly every facet of this study, beginning with its fundamental parameters. The surface gravity of this star (log(g) = 4.366) is much lower than all other members of ABD, and its radius is 1.12 R$_{\odot}$ while its mass is slightly lower (1.03 M$_{\odot}$), indicative of a younger star. This conjecture is supported by its placement on the HR diagram, as seen in Figure \ref{Iso_Paper}, as being nearly 60 Myr. In addition, the chemical composition of this star is much lower than the rest of our ABD sample. While this star does traceback with ABD, the other aspects of this study warrant a younger, more metal poor moving group, and thus we do not consider BD-04 1063 as a member of the ABD moving group.

It is also important to comment on the distance to this star. As mentioned in Section 3.2 and Section 5, this star does not have a Hipparcos parallax estimate, therefore the distance is from Torres et al. (2008). This distance estimate assumes that BD-04 1063 is a member of ABD. With this vital criteria in question, we found a previous distance estimate to this star which uses photometric data to calculate a distance of 61 pc (Cutispoto et al. 2001). This is a much closer distance than the 78 pc given in Torres et al. (2008) and when applied, changes the radius and mass estimates to 0.83 R$_{\odot}$ and 0.58 M$_{\odot}$ respectively. The closer distance also lowers the luminosity to 0.59 L$_{\odot}$ and allows the star to fall along the same isochrone as the rest of ABD. 

\emph{BD-15 200}: The Ba13 study also analyzed this star using high resolution spectroscopy and found nearly identical fundamental parameters and chemical composition. This star has a very similar surface gravity, radius, mass, and lithium abundance to the rest of the moving group, however, its chemical composition is slightly askew. It has a much larger metallicity ([M/H] = 0.03) as compared to the rest of the cluster. In addition, this star diverges from the ABD orbit around 100 Myr which could be a sign that this is a younger, more enriched star associated with the same molecular cloud.

\emph{BD+23 296 A}: This star is the most enigmatic of our sample. It has a relatively close K4 binary companion with a separation of $1\farcs8$ and is labeled in Simbad as a giant type star, even though its temperature coupled with membership in ABD should place it on the ZAMS. Casagrande et al. (2011) computed the age of this star using main sequence evolutionary models and found an age of 13.8 Gyr. The giant status is seen in Figure \ref{Iso_Paper} where the star is lifted off of the MS, making it appear much younger. Apart from its position on the HRD, its surface gravity, chemical composition, and traceback are consistent with other ABD members. We therefore believe this elevation is due to its K4 companion or an unresolved companion adding flux, an incorrect parallax, or some combination of the two.

\emph{BD+41 4749}: This star follows nearly every trend in ABD except the KTb where it deviates quite drastically from the rest of the group. This is likely attributed to its low U velocity and low RV (-19 km s$^{-1}$), causing the epicyclic frequency to be different from other stars in the subsample. Due to its similarities with other members in age, chemical composition, and mass and radius, we are hesitant to demote this star from membership, though we note its peculiar space motion and origin.

\emph{BD-09 1034}: Like BD-04 1063, this star does not have a Hipparcos distance estimate and instead we use the Torres et al. (2008) distance which assumes membership in ABD. Membership appears to be a valid assumption since the stars surface gravity and chemical abundance, tests which are independent of distance, match with the majority of our sample. To ensure this distance estimate is reasonable, we estimated the distance to this star by assuming its radius is the same as 3 other ABD stars in the sample with temperatures $\sim$5550 K (BD+41 4749, IS Eri, and V577 Per A). Using the average radius of 0.80 R$_{\odot}$, we calculated the luminosity and M$_{V}$, added the BC$_{V}$, and found that for V = 9.98, the distance is 78 pc. This is slightly closer than the Torres et al. (2008) estimate of 88 pc. The corresponding mass estimate is 0.70 M$_{\odot}$.

\emph{Comparison to Ba13}: This work and the Ba13 study are complimentary to one another. Only one star overlaps both samples, yet many of the same results for the chemical analysis and kinematics were found in both studies, including nearly identical metal abundances for each ABD sample ([M/H] = 0.01). In addition to the large scale results, several smaller trends exist between the two studies. First, in Ba13 there are 5 stars with [Fe/H] = -0.04 or -0.03 and 3 stars with [Fe/H] = 0.02, 0.05, and 0.06 (with [Fe/H] = 0.06 corresponding to BD-15 200). Apart from probable systematic differences, this matches the 8 stars in our sample with [Fe/H] between -0.01 and 0.01 and 1 star with [Fe/H] = 0.07. Therefore, in Table \ref{ChemBa13} we build a sample of chemically coherent stars in ABD using both studies as well as highlight the outliers. In addition, BD-03 4778 from the Ba13 study has a surface gravity and chemical composition (log(g) = 4.31, [Fe/H] = -0.09) which looks similar to BD-04 1063 (log(g) = 4.37, [Fe/H] = -0.08). This could be an indicator that BD-03 4778 is also not truly a member of ABD.

\section{Conclusion and Future Work}

We have investigated 10 proposed members of the ABD stream to identify bulk characteristics of the moving group so that in the future, these will be used in conjunction with other techniques such as x-ray emission, Li depletion, and H$\alpha$ emission to classify whether future proposed members belong in the moving group. Using TGVIT and MOOG together, these stars have precise spectroscopic parameters (T$_{eff}$, log(g), [Fe/H], $v_{t}$) along with radii, spectroscopic masses, and chemical abundances. We also find that $<[M/H]>$ = -0.03 $\pm$ 0.06, consistent with previous results (Biazzo et al. 2012, Ba13), suggesting this is characteristic of ABD as a whole. Along with other recent age estimates, our results verify that ABD falls along the 100 Myr isochrone and traces back to 125 Myr. After investigating three different evolutionary models, we have found the YREC models to best fit our observational data. 

Our method has shown BD-04 1063 is most likely not a member of the moving group based on surface gravity, chemical composition, and position on the HR Diagram. Our findings also confirm the Ba13 results for the fundamental parameters and chemical composition of BD-15 200 (HD 6569), and it is possible this is a younger star in the ABD moving group. Finally, BD+23 296 A has an odd placement on the HRD which is likely due to a companion (either known or unresolved) or incorrect parallax estimate.
 
The overall procedure presented in the work is a valuable tool for analyzing other, younger moving groups. However, to truly utilize this procedure, we will compile another line list which allows the investigation of faster rotating stars (vsin(i) $\sim$ 30 km s$^{-1}$) as younger groups have not had enough time to spin down. In fact, Eggenberger et al. (2012) showed that once the debris disk evaporate from a star, angular momentum from the still collapsing core is able to spin up the surface of the star. This effect makes the youngest available moving groups (e.g. TW Hydrae and $\beta$ Pictoris) more difficult to analyze as they have recently left their disk-locking phase and have large vsin(i). We therefore find it important to future characterization of younger moving groups to find a robust line list capable of this analysis.

These young moving groups are pivotal for constraining PMS evolutionary models. As mentioned briefly in Section 6, there is large variance in ages and masses between PMS evolutionary models and with upwards of 10 models available, large inconsistencies arise in the literature. We are currently observing many nearby clusters with varying ages to begin constraining the models in the F, G, and early K spectral regime. This type of constraint can be achieved using the procedure outlined in Sections 3.1 and 3.2 of this work by finding spectroscopic masses for several stars in a gambit of moving groups and observing how stars of the same mass evolve as a function of age. ABD already places constraints on where the main sequence should begin, however, this is only a small part of a larger picture.

\acknowledgments

We are grateful for the observational assistance provided by David Doss
at the McDonald Observatory. We would also like to thank Josh Schlieder for 
constructive criticism and ideas and Jennifer Johnson for her expertise in MOOG.
We also thank the anonymous referee for a thorough read of the manuscript as well 
as suggestions which improved the paper.

\appendix
\section{Appendix A}

In order to calculate the covariance between the stellar parameters, we used the bootstrap method outlined in Section 3.1 and found the covariant terms using:

\begin{equation}
	\sigma_{T,logg} = \frac{1}{N}\sum^{N}_{i=1}(T_{i} - \overline{T})(logg_{i} - \overline{logg})
\end{equation}

where $\sigma_{T,logg}$ is the covariance between T and logg. We then follow the outline of Johnson et al. (2002), particularly equation (2), to find total errors in the parameters. For temperature and mictroturbulence, it was necessary to find the logarithmic uncertainty and convert back to the original units to avoid scaling effects. A sample equation is

\begin{equation}
\begin{split}
\sigma_{T_{Total}}^{2} = \sigma_{rand,T}^{2} + [(\frac{\delta logT}{\delta logg})(\frac{\delta logT}{\delta logv_{t}})\sigma_{logg,logv_{t}} \\
 + (\frac{\delta logT}{\delta logg})(\frac{\delta logT}{\delta logA(Fe)})\sigma_{logg,logA(Fe)} \\
 + (\frac{\delta logT}{\delta logv_{t}})(\frac{\delta logT}{\delta logA(Fe)})\sigma_{logv_{t},logA(Fe)}]
\end{split}
\end{equation}

where $\sigma_{rand,T}^{2}$ is the log form or the errors assigned in Table 1. For the star BD+23 296 A, the random temperature errors are $\pm$ 32 K which corresponds to a logarithmic temperature error of 2.624 * 10$^{-3}$ dex. The total covariant error sums up to 3.447 * 10$^{-6}$ and only effects the uncertainty by 0.04 K and we therefore deem this as insignificant. The largest covariant uncertainty was found in logg with an error of 0.00235 dex, still below the threshold to make any significant contribution to the random error. Therefore, TGVIT produces highly uncorrelated parameters since it varies temperature and surface gravity simultaneously.

\clearpage

\clearpage

%\begin{landscape}
%\end{landscape}
%\clearpage

%\begin{landscape}
%\end{landscape}
\clearpage

%\begin{landscape}
%\end{landscape}
\clearpage

%\end{landscape}
\clearpage

\clearpage

\begin{landscape}
\begin{deluxetable}{lrrrrrrrrrr}
  \tabletypesize{\scriptsize}
  \tablewidth{0pt}
  \tablecaption{Parameters for Proposed Moving Group Members and VF05 Stars\label{Param}}
  \tablehead{
    \multicolumn{2}{c}{} &
    \colhead{$T_\textrm{eff}$} & 
    \colhead{log(g)} & 
    \colhead{$v_{t}$} &
    \colhead{[Fe/H]} & 
    \colhead{Radius} & 
    \colhead{Mass} & 
    \colhead{RV} & 
    \colhead{vsin(i)} & 
    \colhead{EW$_{Li}^{a}$}  \\
    \colhead{HD} & 
    \colhead{Other} &
    \colhead{(K)} & 
    \colhead{(dex)} &
    \colhead{(dex)} & 
    \colhead{(km s$^{-1}$)} &               
    \colhead{($R_{\odot}$)} & 
    \colhead{($M_{\odot}$)} &  
    \colhead{(km s$^{-1}$)} & 
    \colhead{(km s$^{-1}$)} & 
    \colhead{(m\AA)}   }  \startdata
		
	\multicolumn{11}{c}{AB Dor Sample}\\
6569 & BD-15 200 & 5157$\pm$36 & 4.617$\pm$0.09 & 1.24$\pm$0.13 & 0.07$\pm$0.03 & 0.75$\pm$0.02 & 0.85$\pm$0.08 & 8.0$\pm$0.6 & 4.4$\pm$1.1 & 142 \\
\nodata & BD-12 243 & 5367$\pm$25 & 4.655$\pm$0.06 & 1.25$\pm$0.11 & 0.00$\pm$0.02 & 0.81$\pm$0.01 & 1.08$\pm$0.06 & 11.2$\pm$0.7 & 3.0$\pm$2.4 & 144 \\
13482 & BD+23 296 A & 5353$\pm$32 & 4.583$\pm$0.08 & 1.33$\pm$0.13 & 0.00$\pm$0.03 & 1.06$\pm$0.02 & 1.58$\pm$0.13 & 0.6$\pm$0.6 & 6.8$\pm$0.8 & 110 \\
16760 & BD+37 604 Aa & 5614$\pm$23 & 4.503$\pm$0.05 & 1.04$\pm$0.09 & 0.00$\pm$0.02 & 0.88$\pm$0.03 & 0.90$\pm$0.04 & -2.4$\pm$0.5 & 1.5$\pm$1.8 & 157 \\
19668 & IS Eri & 5561$\pm$30 & 4.653$\pm$0.07 & 1.44$\pm$0.13 & 0.01$\pm$0.03 & 0.76$\pm$0.01 & 0.94$\pm$0.07 & 14.9$\pm$0.7 & 6.4$\pm$1.6 & 175 \\
\nodata & BD+21 418 A & 5900$\pm$41 & 4.588$\pm$0.08 & 1.69$\pm$0.16 & 0.00$\pm$0.04 & 0.91$\pm$0.04 & 1.17$\pm$0.09 & 6.3$\pm$0.8 & 7.6$\pm$1.1 & 150 \\ 
21845 & V577 Per A & 5552$\pm$34 & 4.536$\pm$0.08 & 1.69$\pm$0.11 & -0.01$\pm$0.03 & 0.79$\pm$0.01 & 0.77$\pm$0.06 & -4.7$\pm$0.8 & 8.9$\pm$0.9 & 215 \\
31652 & BD-09 1034 & 5553$\pm$45 & 4.477$\pm$0.12 & 1.70$\pm$0.18 & -0.01$\pm$0.04 & 0.90$\pm$0.06 & 0.89$\pm$0.11 & 23.6$\pm$0.7 & 5.7$\pm$1.0 & 242 \\
293857 & BD-04 1063 & 5572$\pm$44 & 4.366$\pm$0.10 & 1.95$\pm$0.20 & -0.08$\pm$0.04 & 1.12$\pm$0.06 & 1.03$\pm$0.10 & 21.7$\pm$0.8 & 8.6$\pm$1.0 & 240 \\
\nodata & BD+41 4749 & 5532$\pm$22 & 4.575$\pm$0.05 & 1.33$\pm$0.10 & -0.01$\pm$0.02 & 0.84$\pm$0.01 & 0.97$\pm$0.03 & -19.2$\pm$0.6 & 4.4$\pm$0.8 & 163 \\
\hline
	\multicolumn{11}{c}{VF05 Sample}\\
42618 & BD+06 1155 & 5735$\pm$18 & 4.478$\pm$0.04 & 1.05$\pm$0.09 & -0.10$\pm$0.02 & 0.93$\pm$0.01 & 0.94$\pm$0.04 & -52.3$\pm$0.4 & 4.4$\pm$0.9 & \nodata \\
47127 & BD+12 1219 & 5608$\pm$23 & 4.341$\pm$0.06 & 1.05$\pm$0.08 & 0.12$\pm$0.02 & 1.15$\pm$0.01 & 1.06$\pm$0.06 & 50.7$\pm$0.5 & 3.5$\pm$1.4 & \nodata \\
65583 & BD+29 1664 & 5265$\pm$24 & 4.507$\pm$0.06 & 0.76$\pm$0.16 & -0.71$\pm$0.02 & 0.79$\pm$0.01 & 0.73$\pm$0.04 & 15.8$\pm$0.5 & 3.3$\pm$1.7 & \nodata \\
68017 & STT 564 A & 5563$\pm$22 & 4.459$\pm$0.06 & 0.89$\pm$0.13 & -0.41$\pm$0.02 & 0.97$\pm$0.01 & 0.98$\pm$0.06 & 30.6$\pm$0.6 & 1.5$\pm$1.6 & \nodata \\
73668 & BD+06 2007 & 5928$\pm$24 & 4.391$\pm$0.06 & 1.18$\pm$0.1 & 0.01$\pm$0.02 & 1.12$\pm$0.02 & 1.12$\pm$0.07 & -19.8$\pm$0.6 & 4.5$\pm$1.4 & 48.7 \\
77407 & MUG 1 & 5931$\pm$28 & 4.473$\pm$0.07 & 1.34$\pm$0.12 & 0.05$\pm$0.03 & 1.02$\pm$0.01 & 1.14$\pm$0.08 & 5.8$\pm$0.7 & 7.4$\pm$1.1 & 163 \\
100167 & BD+42 2216 & 5925$\pm$23 & 4.475$\pm$0.06 & 1.09$\pm$0.1 & 0.09$\pm$0.02 & 1.03$\pm$0.01 & 1.15$\pm$0.08 & -3.6$\pm$0.6 & 5.0$\pm$1.0 & 104 \\
101501 & 61 UMa & 5527$\pm$20 & 4.612$\pm$0.05 & 1.08$\pm$0.1 & -0.04$\pm$0.02 & 0.86$\pm$0.01 & 1.09$\pm$0.05 & -4.5$\pm$0.4 & 1.9$\pm$1.5 & \nodata \\
111395 & LW Com & 5620$\pm$23 & 4.543$\pm$0.05 & 1.13$\pm$0.1 & 0.08$\pm$0.02 & 0.92$\pm$0.01 & 1.08$\pm$0.04 & -7.8$\pm$0.6 & 3.8$\pm$0.8 & 24.3 \\
111398 & BD+12 2518 & 5716$\pm$21 & 4.275$\pm$0.05 & 1.09$\pm$0.08 & 0.10$\pm$0.02 & 1.36$\pm$0.01 & 1.26$\pm$0.06 & 4.0$\pm$0.6 & 4.2$\pm$1.0 & \nodata \\
117176 & 70 Vir & 5482$\pm$22 & 3.946$\pm$0.06 & 1.1$\pm$0.07 & -0.08$\pm$0.02 & 1.91$\pm$0.02 & 1.17$\pm$0.07 & 5.7$\pm$0.6 & 5.0$\pm$1.2 & 32.4 \\
121320 & BD+21 2588 & 5568$\pm$19 & 4.593$\pm$0.05 & 0.94$\pm$0.11 & -0.21$\pm$0.02 & 0.85$\pm$0.01 & 1.02$\pm$0.05 & -11.2$\pm$0.5 & 2.3$\pm$1.3 & \nodata \\
\hline
\nodata & Sun & 5763$\pm$20 & 4.463$\pm$.05 &  1.06$\pm$0.08  &0.00$\pm$.02 & \nodata & \nodata & 0.4$\pm$0.5 & 0.6$\pm$0.8 & \nodata\\

%\nodata & BD+41 4749 & 5519$\pm$23 & 4.567$\pm$0.050 & -0.05$\pm$0.03 & 1.32$\pm$0.11 & 0.85 & 0.97 & -19.27 & 6.5 & 2.82\\
%\nodata & BD+21 418 A & 5823$\pm$30 & 4.487$\pm$0.070 & -0.08$\pm$0.03 & 1.68$\pm$0.10 & 0.93 & 0.98 & 6.26 & 9.4 & 3.03\\
%293857 & BD-04 1063 & 5489$\pm$33 & 4.067$\pm$0.075 & -0.21$\pm$0.04 & 2.20$\pm$0.17 & 1.16 & 0.57 & 21.57 & 8.5 & 3.27\\
%16760 & BD+37 604 Aa & 5636$\pm$18 & 4.515$\pm$0.035 & 0.00$\pm$0.02 & 1.00$\pm$0.09 & 0.87 & 0.91 & -2.51 & 2.5 & 2.89\\
%6569 & BD-15 200 & 5155$\pm$50 & 4.621$\pm$0.100 & 0.04$\pm$0.03 & 1.29$\pm$0.13 & 0.75 & 0.86 & 7.89 & 4.3 & 2.26\\
%19668 & IS Eri & 5545$\pm$30 & 4.592$\pm$0.065 & -0.03$\pm$0.03 & 1.43$\pm$0.12 & 0.76 & 0.82 & 14.84 & 7.4 & 2.89\\
%13482 & BD+23 296 A & 5401$\pm$25 & 4.640$\pm$0.055 & -0.01$\pm$0.03 & 1.39$\pm$0.13 & 1.04 & 1.71 & 0.59 & 6.8 & 2.38\\
%31652 & BD-09 1034 & 5414$\pm$45 & 4.227$\pm$0.075 & -0.18$\pm$0.03 & 1.95$\pm$0.14 & 0.96 & 0.57 & 23.48 & 7.0 & 3.22\\
%\nodata & BD-12 243 & 5381$\pm$28 & 4.554$\pm$0.07 & -0.04$\pm$0.03 & 1.37$\pm$0.10 & 0.80 & 0.83 & 11.08 & 3.4 & 2.54\\
%21845 & V577 Per A & 5527$\pm$40 & 4.536$\pm$0.09 & -0.06$\pm$0.04 & 1.75$\pm$0.13 & 0.79 & 0.78 & -4.89 & 9.1 & 3.13\\
    \enddata
    \tablecomments{
		$^{a}$ Errors in EW$_{Li}$ are 5\%
}
\end{deluxetable}
\end{landscape}
\clearpage

\begin{landscape}
\begin{deluxetable}{lrrrrrrr} \tabletypesize{\scriptsize}
  \tablewidth{0pt}
  \tablecaption{Chemical Abundances of ABD and VF05 Stars\label{Abun}}
  \tablehead{
    \colhead{Star} & 
    \colhead{[Na/H]}   & 
    \colhead{[Mg/H]}   &
    \colhead{[Al/H]}   & 
    \colhead{[Si/H]}   & 
    \colhead{[Ca/H]}   & 
	\colhead{[ScI/H]}  &
	\colhead{[ScII/H]}  \\
		\colhead{(A(X)$_{Sun}$)} &
	\colhead{(6.39)}  &
	\colhead{(7.62)}  &
	\colhead{(6.44)}  &
	\colhead{(7.54)}  &
	\colhead{(6.36)}  &
	\colhead{(3.11)}  &
	\colhead{(3.06)}  
		
	}  \startdata
	\multicolumn{8}{c}{AB Dor Sample} \\
    BD-15 200 & -0.06 $\pm$ 0.12 & -0.11 $\pm$ 0.10 &  0.07 $\pm$ 0.05 &  0.03 $\pm$ 0.06 &  0.05 $\pm$ 0.10 &  0.02 $\pm$ 0.13 & -0.06 $\pm$ 0.11 \\
    BD-12 243 & -0.02 $\pm$ 0.05 & -0.07 $\pm$ 0.06 & -0.04 $\pm$ 0.03 & -0.01 $\pm$ 0.03 &  0.03 $\pm$ 0.07 & -0.06 $\pm$ 0.09 & -0.06 $\pm$ 0.07 \\
  BD+23 296 A & -0.02 $\pm$ 0.06 & -0.11 $\pm$ 0.06 & -0.03 $\pm$ 0.04 & -0.06 $\pm$ 0.04 &  0.04 $\pm$ 0.09 & -0.07 $\pm$ 0.14 & -0.11 $\pm$ 0.09 \\
 BD+37 604 Aa & -0.07 $\pm$ 0.04 &  0.04 $\pm$ 0.09 & -0.01 $\pm$ 0.03 & -0.02 $\pm$ 0.02 &  0.03 $\pm$ 0.05 & -0.12 $\pm$ 0.04 & -0.06 $\pm$ 0.05 \\
       IS Eri & -0.06 $\pm$ 0.06 & -0.05 $\pm$ 0.08 & -0.07 $\pm$ 0.03 & -0.05 $\pm$ 0.03 & -0.02 $\pm$ 0.08 &  0.03 $\pm$ 0.07 & -0.10 $\pm$ 0.09 \\
  BD+21 418 A & -0.04 $\pm$ 0.06 &  0.00 $\pm$ 0.07 & -0.06 $\pm$ 0.04 & -0.05 $\pm$ 0.03 &  0.05 $\pm$ 0.10 &     \nodata      & -0.15 $\pm$ 0.10  \\
   V577 Per A & -0.03 $\pm$ 0.06 & -0.05 $\pm$ 0.10 & -0.04 $\pm$ 0.10 & -0.03 $\pm$ 0.03 &  0.10 $\pm$ 0.09 &     \nodata      &  0.02 $\pm$ 0.13 \\
   BD-09 1034 & -0.03 $\pm$ 0.10 & -0.28 $\pm$ 0.10 &  0.00 $\pm$ 0.10 & -0.16 $\pm$ 0.04 &  0.04 $\pm$ 0.13 & -0.07 $\pm$ 0.10 & -0.16 $\pm$ 0.13 \\
   BD-04 1063 & -0.17 $\pm$ 0.14 & -0.20 $\pm$ 0.08 &  0.02 $\pm$ 0.10 & -0.07 $\pm$ 0.04 &  0.03 $\pm$ 0.13 & -0.02 $\pm$ 0.10 & -0.17 $\pm$ 0.07 \\
   BD+41 4749 & -0.04 $\pm$ 0.04 & -0.01 $\pm$ 0.04 &  0.00 $\pm$ 0.03 & -0.01 $\pm$ 0.02 &  0.05 $\pm$ 0.06 &  0.00 $\pm$ 0.04 & -0.04 $\pm$ 0.06 \\
     Average   &-0.05 $\pm$ 0.04 & -0.08 $\pm$ 0.10 & -0.02 $\pm$ 0.04 & -0.04 $\pm$ 0.05 &  0.04 $\pm$ 0.03 & -0.04 $\pm$ 0.05 & -0.09 $\pm$ 0.06 \\
	 Without BD-04 or BD -15 & -0.04$\pm$0.02 & -0.07$\pm$0.10 & -0.03$\pm$0.03 & -0.05$\pm$0.05 & 0.04$\pm$0.03 & -0.05$\pm$0.05 & -0.08$\pm$0.06  \\
\hline
	\multicolumn{8}{c}{VF05 Sample} \\
     HD 42618 & -0.10 $\pm$ 0.03 & -0.05 $\pm$ 0.03 & -0.07 $\pm$ 0.03 & -0.10 $\pm$ 0.02 & -0.07 $\pm$ 0.04 & -0.18 $\pm$ 0.10 & -0.08 $\pm$ 0.05 \\
     HD 47127 &  0.07 $\pm$ 0.05 &  0.27 $\pm$ 0.05 &  0.23 $\pm$ 0.02 &  0.18 $\pm$ 0.02 &  0.13 $\pm$ 0.06 &  0.11 $\pm$ 0.06 &  0.18 $\pm$ 0.06 \\
     HD 65583 & -0.57 $\pm$ 0.04 & -0.32 $\pm$ 0.07 & -0.41 $\pm$ 0.04 & -0.45 $\pm$ 0.02 & -0.43 $\pm$ 0.07 & -0.49 $\pm$ 0.07 & -0.49 $\pm$ 0.07 \\
     HD 68017 & -0.31 $\pm$ 0.04 & -0.17 $\pm$ 0.04 & -0.10 $\pm$ 0.03 & -0.22 $\pm$ 0.02 & -0.23 $\pm$ 0.06 & -0.21 $\pm$ 0.06 & -0.21 $\pm$ 0.07 \\
     HD 73668 & -0.12 $\pm$ 0.03 & -0.01 $\pm$ 0.04 & -0.04 $\pm$ 0.04 &  0.01 $\pm$ 0.02 &  0.04 $\pm$ 0.06 & -0.03 $\pm$ 0.04 & -0.04 $\pm$ 0.06 \\
     HD 77407 &  0.01 $\pm$ 0.04 & -0.04 $\pm$ 0.06 & -0.12 $\pm$ 0.02 & -0.01 $\pm$ 0.02 &  0.09 $\pm$ 0.07 &     \nodata      &  0.00 $\pm$ 0.08 \\
    HD 100167 & -0.01 $\pm$ 0.04 &  0.07 $\pm$ 0.05 & -0.03 $\pm$ 0.04 &  0.07 $\pm$ 0.02 &  0.10 $\pm$ 0.06 & -0.03 $\pm$ 0.07 &  0.10 $\pm$ 0.08\\
    HD 101501 & -0.07 $\pm$ 0.04 & -0.03 $\pm$ 0.06 & -0.07 $\pm$ 0.02 & -0.05 $\pm$ 0.02 & -0.04 $\pm$ 0.06 & -0.10 $\pm$ 0.06 & -0.06 $\pm$ 0.06 \\
    HD 111395 &  0.06 $\pm$ 0.05 &  0.11 $\pm$ 0.05 &  0.09 $\pm$ 0.02 &  0.08 $\pm$ 0.02 &  0.13 $\pm$ 0.06 &  0.10 $\pm$ 0.05 &  0.05 $\pm$ 0.07 \\
    HD 111398 &  0.04 $\pm$ 0.03 &  0.08 $\pm$ 0.04 &  0.16 $\pm$ 0.03 &  0.13 $\pm$ 0.02 &  0.14 $\pm$ 0.05 &  0.08 $\pm$ 0.04 &  0.21 $\pm$ 0.06 \\
    HD 117176 & -0.11 $\pm$ 0.05 &  0.08 $\pm$ 0.06 &  0.03 $\pm$ 0.03 & -0.02 $\pm$ 0.02 & -0.02 $\pm$ 0.05 &     \nodata      &  0.01 $\pm$ 0.06 \\
    HD 121320 & -0.29 $\pm$ 0.04 & -0.25 $\pm$ 0.05 & -0.25 $\pm$ 0.02 & -0.24 $\pm$ 0.02 & -0.23 $\pm$ 0.05 &     \nodata      & -0.24 $\pm$ 0.06 \\

\enddata
    
\end{deluxetable}
\end{landscape}
\clearpage

\begin{landscape}
\begin{deluxetable}{lrrrrrrrrr}\tabletypesize{\scriptsize}
  \tablewidth{0pt}
  \tablecaption{Chemical Abundances of ABD and VF05 Stars Ctd\label{Ctd}}
  \tablehead{
    \colhead{Star} & 
    \colhead{[TiI/H]}  & 
	\colhead{[TiII/H]} &
    \colhead{[V/H]}    & 
    \colhead{[CrI/H]}  &
	\colhead{[CrII/H]} &
    \colhead{[Mn/H]}   & 
    \colhead{[Co/H]}   &
    \colhead{[Ni/H]}   \\
	\colhead{(A(X)$_{Sun}$)} &
	\colhead{(4.96)}  &
	\colhead{(4.93)}  &
	\colhead{(3.95)}  &
	\colhead{(5.62)}  &
	\colhead{(5.68)}  &
	\colhead{(5.39)}  &
	\colhead{(4.88)}  &
	\colhead{(6.23)}
	}  \startdata

	\multicolumn{9}{c}{AB Dor Sample} \\
    BD-15 200  &  0.09 $\pm$ 0.11 & -0.02 $\pm$ 0.10 &  0.34 $\pm$ 0.12 &  0.04 $\pm$ 0.09 & -0.02 $\pm$ 0.13 &  0.11 $\pm$ 0.07 & -0.02 $\pm$ 0.07 &  0.00 $\pm$ 0.06\\
    BD-12 243  &  0.00 $\pm$ 0.07 & -0.03 $\pm$ 0.06 &  0.14 $\pm$ 0.07 &  0.01 $\pm$ 0.06 & -0.05 $\pm$ 0.12 & -0.05 $\pm$ 0.07 & -0.06 $\pm$ 0.03 & -0.05 $\pm$ 0.04\\
  BD+23 296 A  &  0.00 $\pm$ 0.09 & -0.07 $\pm$ 0.09 &  0.17 $\pm$ 0.11 & -0.03 $\pm$ 0.08 & -0.10 $\pm$ 0.11 & -0.12 $\pm$ 0.07 & -0.12 $\pm$ 0.05 & -0.11 $\pm$ 0.05\\
 BD+37 604 Aa  &  0.02 $\pm$ 0.06 &  0.02 $\pm$ 0.05 &  0.07 $\pm$ 0.06 &  0.05 $\pm$ 0.05 &  0.05 $\pm$ 0.06 & -0.02 $\pm$ 0.05 & -0.04 $\pm$ 0.03 & -0.05 $\pm$ 0.03\\
       IS Eri  &  0.00 $\pm$ 0.08 & -0.07 $\pm$ 0.08 &  0.14 $\pm$ 0.08 & -0.02 $\pm$ 0.07 & -0.17 $\pm$ 0.12 & -0.08 $\pm$ 0.08 & -0.07 $\pm$ 0.05 & -0.08 $\pm$ 0.04\\
  BD+21 418 A  & -0.04 $\pm$ 0.10 & -0.12 $\pm$ 0.10 &  0.14 $\pm$ 0.10 &  0.02 $\pm$ 0.10 & -0.03 $\pm$ 0.15 & -0.12 $\pm$ 0.14 & -0.19 $\pm$ 0.06 & -0.07 $\pm$ 0.05 \\
   V577 Per A  &  0.07 $\pm$ 0.08 & -0.02 $\pm$ 0.09 &  0.15 $\pm$ 0.09 &  0.08 $\pm$ 0.08 &  0.12 $\pm$ 0.10 & -0.06 $\pm$ 0.05 & -0.13 $\pm$ 0.05 & -0.08 $\pm$ 0.05 \\
   BD-09 1034  & -0.05 $\pm$ 0.12 & -0.12 $\pm$ 0.12 &  0.17 $\pm$ 0.11 &  0.01 $\pm$ 0.12 &     \nodata      & -0.21 $\pm$ 0.06 &     \nodata      & -0.16 $\pm$ 0.06 \\
   BD-04 1063  & -0.13 $\pm$ 0.11 & -0.05 $\pm$ 0.07 &  0.07 $\pm$ 0.10 & -0.01 $\pm$ 0.13 & -0.04 $\pm$ 0.11 & -0.18 $\pm$ 0.13 & -0.22 $\pm$ 0.06 & -0.20 $\pm$ 0.06\\
   BD+41 4749  &  0.02 $\pm$ 0.06 &  0.07 $\pm$ 0.07 &  0.12 $\pm$ 0.07 &  0.05 $\pm$ 0.05 &  0.06 $\pm$ 0.08 & -0.05 $\pm$ 0.05 & -0.09 $\pm$ 0.04 & -0.07 $\pm$ 0.03\\
     Average   &  0.00 $\pm$ 0.06 & -0.04 $\pm$ 0.06 &  0.15 $\pm$ 0.08 &  0.02 $\pm$ 0.03 & -0.02 $\pm$ 0.09 & -0.08 $\pm$ 0.09 & -0.10 $\pm$ 0.07 & -0.09 $\pm$ 0.06\\
	 Without BD-04 or BD -15        &  0.00 $\pm$ 0.04 & -0.04 $\pm$ 0.07 & 0.14 $\pm$ 0.03 & 0.02 $\pm$ 0.04 & -0.02 $\pm$ 0.10 & -0.09 $\pm$ 0.06 & -0.10 $\pm$ 0.05 & -0.08 $\pm$ 0.04 \\
	\\
\hline
	\multicolumn{9}{c}{VF05 Sample} \\
     HD 42618  & -0.07 $\pm$ 0.04 & -0.08 $\pm$ 0.04 & -0.14 $\pm$ 0.07 & -0.09 $\pm$ 0.04 &  0.02 $\pm$ 0.08 & -0.15 $\pm$ 0.04 & -0.11 $\pm$ 0.03 & -0.12 $\pm$ 0.03\\
     HD 47127  &  0.16 $\pm$ 0.06 &  0.23 $\pm$ 0.07 &  0.18 $\pm$ 0.05 &  0.12 $\pm$ 0.05 &  0.16 $\pm$ 0.08 &  0.09 $\pm$ 0.06 &  0.14 $\pm$ 0.04 &  0.11 $\pm$ 0.03\\
     HD 65583  & -0.32 $\pm$ 0.08 & -0.35 $\pm$ 0.07 & -0.42 $\pm$ 0.07 & -0.57 $\pm$ 0.06 & -0.53 $\pm$ 0.10 & -0.93 $\pm$ 0.05 & -0.55 $\pm$ 0.04 & -0.64 $\pm$ 0.03\\
     HD 68017  & -0.15 $\pm$ 0.06 & -0.15 $\pm$ 0.07 & -0.21 $\pm$ 0.06 & -0.33 $\pm$ 0.05 & -0.34 $\pm$ 0.07 & -0.64 $\pm$ 0.04 & -0.29 $\pm$ 0.05 & -0.37 $\pm$ 0.03\\
     HD 73668  & -0.01 $\pm$ 0.05 &  0.04 $\pm$ 0.06 & -0.06 $\pm$ 0.06 &  0.00 $\pm$ 0.05 & -0.06 $\pm$ 0.10 & -0.08 $\pm$ 0.04 & -0.08 $\pm$ 0.04 & -0.04 $\pm$ 0.03\\
     HD 77407  & -0.02 $\pm$ 0.06 &  0.03 $\pm$ 0.08 & -0.03 $\pm$ 0.10 &  0.05 $\pm$ 0.06 &  0.09 $\pm$ 0.11 & -0.10 $\pm$ 0.07 & -0.12 $\pm$ 0.04 & -0.06 $\pm$ 0.04 \\
    HD 100167  &  0.04 $\pm$ 0.05 &  0.06 $\pm$ 0.07 &  0.03 $\pm$ 0.04 &  0.09 $\pm$ 0.04 &  0.13 $\pm$ 0.09 & -0.08 $\pm$ 0.06 &  0.01 $\pm$ 0.05 &  0.03 $\pm$ 0.03\\
    HD 101501  & -0.04 $\pm$ 0.06 &  0.00 $\pm$ 0.06 &  0.03 $\pm$ 0.05 &  0.01 $\pm$ 0.05 &  0.01 $\pm$ 0.09 & -0.08 $\pm$ 0.05 & -0.06 $\pm$ 0.03 & -0.06 $\pm$ 0.03\\
    HD 111395  &  0.09 $\pm$ 0.06 &  0.11 $\pm$ 0.06 &  0.11 $\pm$ 0.06 &  0.10 $\pm$ 0.05 &  0.11 $\pm$ 0.10 &  0.11 $\pm$ 0.05 &  0.04 $\pm$ 0.04 &  0.04 $\pm$ 0.04\\
    HD 111398  &  0.13 $\pm$ 0.05 &  0.20 $\pm$ 0.07 &  0.11 $\pm$ 0.05 &  0.12 $\pm$ 0.05 &  0.18 $\pm$ 0.08 &  0.10 $\pm$ 0.04 &  0.09 $\pm$ 0.04 &  0.09 $\pm$ 0.03\\
    HD 117176  & -0.02 $\pm$ 0.06 &  0.04 $\pm$ 0.06 & -0.06 $\pm$ 0.06 & -0.06 $\pm$ 0.05 &  0.02 $\pm$ 0.07 & -0.18 $\pm$ 0.05 & -0.09 $\pm$ 0.04 & -0.09 $\pm$ 0.03\\
    HD 121320  & -0.22 $\pm$ 0.05 & -0.20 $\pm$ 0.06 & -0.22 $\pm$ 0.05 & -0.23 $\pm$ 0.05 & -0.16 $\pm$ 0.07 & -0.38 $\pm$ 0.06 & -0.32 $\pm$ 0.04 & -0.30 $\pm$ 0.03\\

\enddata
    
\end{deluxetable}
\end{landscape}
\clearpage

\begin{deluxetable}{ll}
  \tabletypesize{\scriptsize}
  \tablewidth{0pt}
  \tablecaption{Galactic Potential Parameters \label{Gal}}
  \tablehead{
    \colhead{Parameters} & 
    \colhead{Value}     }  \startdata
				R$_{\odot}$ & 8 kpc \\
				$\omega_{\odot}$ & 227.7 km s$^{-1}$ \\
				$\rho_{\odot}$ & 0.1 M$_{\odot}$ pc$^{-3}$ \\
				%\hline
				M$_{Disk}$ & 10$^{11}$ M$_{\odot}$ \\
				M$_{Bulge}$ & 3.4 $\bullet$ 10$^{10}$ M$_{\odot}$ \\
				a$_{Disk}$ & 6.5 kpc \\
				b$_{Disk}$ & 0.26 kpc \\
				c$_{Bulge}$ & 0.7 kpc \\
				d$_{Halo}$ & 12.0 kpc \\
				V$_{Halo}$ & 128 km s$^{-1}$ \\
\enddata
    
\end{deluxetable}
\clearpage

\begin{deluxetable}{lrrrrrr}
  \tabletypesize{\scriptsize}
  \tablewidth{0pt}
  \tablecaption{Space Motions and Positions\label{UVW}}
  \tablehead{
    \colhead{Star} & 
    \colhead{$\xi_{i}'$} & 
    \colhead{$\eta_{i}'$} & 
    \colhead{$\zeta_{i}'$} &
    \colhead{U} & 
    \colhead{V} &
    \colhead{W}   \\
    \colhead{\nodata} & 
    \colhead{(pc)} &
    \colhead{(pc)} &
    \colhead{(pc)} & 
    \colhead{(km s$^{-1}$)} &
    \colhead{(km s$^{-1}$)} & 
    \colhead{(km s$^{-1}$)}    }  \startdata
BD-15 200 & 7.6$\pm$0.3  &  -8.5$\pm$0.3  &  -46.1$\pm$1.5  &  -8.2$\pm$0.3  &  -28.7$\pm$1.0  &  -11.3$\pm$0.6\\
BD-12 243 & 5.8$\pm$0.1  &  -8.9$\pm$0.2  &  -32.7$\pm$0.7  &  -5.1$\pm$0.2  &  -26.6$\pm$0.6  &  -15.0$\pm$0.6\\
BD+23 296 A & 16.9$\pm$0.4  &  -24.7$\pm$0.6  &  -21.2$\pm$0.5  &  -9.3$\pm$0.5  &  -31.0$\pm$0.9  &  -14.9$\pm$0.5\\
BD+37 604 A & 24.4$\pm$1.5  &  -35.5$\pm$2.2  &  -15.1$\pm$1.0  &  -8.9$\pm$0.8  &  -25.0$\pm$1.5  &  -11.8$\pm$0.9\\
IS Eri & -3.8$\pm$0.1  &  -22.5$\pm$0.5  &  -29.6$\pm$0.7  &  -6.3$\pm$0.4  &  -27.0$\pm$0.6  &  -10.5$\pm$0.5\\
BD+21 418 A & 15.4$\pm$0.6  &  -44.0$\pm$1.6  &  -26.9$\pm$1.0  &  -6.7$\pm$0.5  &  -29.5$\pm$1.1  &  -18.6$\pm$0.7\\
V577Per A & 17.1$\pm$0.4  &  -29.4$\pm$0.6  &  -4.8$\pm$0.1  &  -6.8$\pm$0.5  &  -26.0$\pm$0.6  &  -16.1$\pm$0.4\\
BD-09 1034 & -35.9$\pm$2.3  &  -68.0$\pm$4.4  &  -43.2$\pm$2.8  &  -6.7$\pm$0.9  &  -26.9$\pm$1.2  &  -15.0$\pm$0.6\\
BD-04 1063 & -20.3$\pm$0.8  &  -44.5$\pm$1.7  &  -21.8$\pm$0.8  &  -8.4$\pm$0.6  &  -18.4$\pm$0.5  &  -18.8$\pm$0.5\\
BD+41 4749 & 46.3$\pm$1.5  &  -12.7$\pm$0.4  &  -15.1$\pm$0.5  &  -4.4$\pm$0.4  &  -27.0$\pm$0.6  &  -14.8$\pm$0.7\\
\\
Average & 7.3$\pm$23.1 & -29.9$\pm$18.8 & -25.7$\pm$12.9 & -7.1$\pm$1.6 & -26.6$\pm$3.4 & -14.7$\pm$2.8\\
\enddata 
    \tablecomments{
$\xi_{i}'$, $\eta_{i}'$, and $\zeta_{i}'$ are the present X, Y, and Z coordinates of the stars.
}
\end{deluxetable}
\clearpage

\begin{deluxetable}{lrrrr}
  \tabletypesize{\scriptsize}
  \tablewidth{0pt}
  \tablecaption{Chemically Consistent Members From This Work and Ba13\label{ChemBa13}}
  \tablehead{
    \colhead{Star} & 
    \colhead{T$_{eff}$} & 
    \colhead{log(g)} & 
    \colhead{[Fe/H]} &
    \colhead{v$_{t}$}   \\   }  \startdata
		\multicolumn{5}{c}{Smaller [Fe/H] Sample} \\
BD-12 243 & 5367 & 4.655 & 0.00 & 1.25 \\ 
BD+23 296 A & 5353 & 4.583 & 0.00 & 1.33 \\
BD+37 604 Aa & 5614 & 4.503 & 0.00 & 1.04 \\
IS Eri & 5561 & 4.653 & 0.01 & 1.44 \\
BD+21 418 A & 5900 & 4.588 & 0.00 & 1.69 \\
V577 Per A & 5552 & 4.536 & -0.01 & 1.69 \\
BD-09 1034 & 5553 & 4.477 & -0.01 & 1.7 \\
HD 317617$^{Ba13}$ & 4870 & 4.49 & -0.03 & 1.10 \\ 
HD 189285$^{Ba13}$ & 5537 & 4.46 & -0.03 & 1.51 \\
HD 199058$^{Ba13}$ & 5737 & 4.62 & -0.03 & 1.05 \\
HD 207278$^{Ba13}$ & 5710 & 4.56 & 0.02 & 1.70 \\
HD 217343$^{Ba13}$ & 5830 & 4.59 & -0.04 & 1.70 \\
BD+41 4749 & 5532 & 4.575 & -0.01 & 1.33 \\
HD 224228$^{Ba13}$ & 4953 & 4.56 & -0.04 & 1.11 \\
\hline
\multicolumn{5}{c}{Larger [Fe/H] Sample} \\
BD-15 200 & 5157 & 4.617 & 0.07 & 1.24 \\
HD 6569$^{Ba13,a}$ & 5170 & 4.61 & 0.06 & 1.37 \\
HD 218860A$^{Ba13}$ & 5543 & 4.59 & 0.05 & 1.45 \\
\hline
\multicolumn{5}{c}{Outliers} \\
BD-04 1063 & 5572 & 4.366 & -0.08 & 1.95 \\
BD-03 4778$^{Ba13}$ & 5220 & 4.31 & -0.09 & 1.80 \\ 
TYC 486-4943-1$^{Ba13}$ & 5160 & 4.87 & -0.10 & 2.50 \\
\\
\enddata 
    \tablecomments{
$^{Ba13}$ From the Ba13 study.
$^{a}$ Same star as BD-15 200.
A Comprehensive list of all stars from this study and Ba13 showing the two [Fe/H] samples in ABD as well as the outlying stars.
}
\end{deluxetable}
\clearpage

\clearpage

\end{document}